\documentclass[aps,prd,reprint,longbibliography,nofootinbib,superscriptaddress]{revtex4-2}

\usepackage{slashed}
\usepackage{verbatim}
\usepackage[T1]{fontenc}
\usepackage{mathbbol}
\usepackage[dvipsnames]{xcolor}
\usepackage{orcidlink}
\usepackage[english]{babel}
\usepackage{lipsum}
\usepackage{physics}
\usepackage{array}
\usepackage{dcolumn}
\usepackage{url}
\usepackage{tensor}
\usepackage{comment}
\usepackage{graphicx,color,overpic,mathtools} 
\usepackage{amsthm,amsmath,amssymb,mathrsfs}
\usepackage{braket,bm,bbm,setspace}
\usepackage{cancel}
\usepackage{float}
\usepackage{xargs}
\definecolor{teal}{RGB}{0, 128, 128}
\definecolor{myred}{RGB}{179, 27, 27}
\usepackage{hyperref}
\hypersetup{colorlinks=true, linkcolor=blue, citecolor=green}

\definecolor{red}{rgb}{1,0,0}

\def\+{^\dagger}

\def\<{\leftarrow}
\def\>{\rightarrow}

\def\({\left(}
\def\){\right)}

\newcommand{\bi}{\begin{itemize}} 				\newcommand{\ei}{\end{itemize}}
\newcommand{\benu}{\begin{enumerate}} 		\newcommand{\enu}{\end{enumerate}}
\newcommand{\bd}{\begin{dinglist}{0}}     \newcommand{\ed}{\end{dinglist}}
\newcommand{\bfig}{\begin{figure}[htbp]}  \newcommand{\efig}{\end{figure}}
        			
\newcommand{\bc}{\begin{center}} 				  \newcommand{\ec}{\end{center}}
\newcommand{\be}{\begin{equation}} 				\newcommand{\ee}{\end{equation}}
\newcommand{\bsub}{\begin{subequations}}  \newcommand{\esub}{\end{subequations}}
\newcommand{\ben}{\begin{eqnarray}} 			\newcommand{\een}{\end{eqnarray}}
\newcommand{\ba}[1]{\begin{array}{#1}} 		\newcommand{\ea}{\end{array}}
\newcommand{\bea}{\begin{equation}\begin{array}{rcl}}
\newcommand{\eea}{\end{array}\end{equation}}

\begin{document}
\title{Multi-photon ring structure of  reflection-asymmetric traversable thin-shell wormholes}

\author{Caio F. B. Macedo}
\email{caiomacedo@ufpa.br}
\affiliation{Faculdade de F\'{i}sica, Campus Salin\'{o}polis, Universidade Federal do Par\'{a}, 68721-000, Salin\'{o}polis, Par\'{a}, Brazil}
\affiliation{		Programa de P\'{o}s-Gradua\c{c}\~{a}o em F\'{i}sica, Universidade Federal do Par\'{a}, 66075-110, Bel\'{e}m, PA, Brazil}

\author{Jo\~{a}o Lu\'{i}s Rosa}
\email{joaoluis92@gmail.com}
\affiliation{Departamento de F\'isica Te\'orica, Universidad Complutense de Madrid, E-28040 Madrid, Spain}
\affiliation{Institute of Physics, University of Tartu, W. Ostwaldi 1, 50411 Tartu, Estonia}

\author{Diego Rubiera-Garcia} \email{drubiera@ucm.es}
\affiliation{Departamento de F\'isica Te\'orica and IPARCOS,
	Universidad Complutense de Madrid, E-28040 Madrid, Spain}
\author{Alejandro Rueda} \email{alerue02@ucm.es}
\affiliation{Departamento de F\'isica Te\'orica and IPARCOS,
	Universidad Complutense de Madrid, E-28040 Madrid, Spain}

\date{\today}
\begin{abstract}
We consider the observational signatures of thin accretion disks around a reflection-asymmetric traversable thin-shell wormhole. This wormhole, built in the framework of Palatini $f(R)$ gravity coupled to a Maxwell field using a junction conditions formalism, lacks horizons but features photon spheres on each side of the throat, described by different effective potentials and at different locations. This fact allows a portion of the light rays arriving to the observer's screen on one side of the throat to have explored a part of the space-time on the other side, bringing information about the geometry gathered there. In this setting we simulate the optical appearance of such an asymmetric wormhole when illuminated by thin accretion disks, investigating scenarios with either one or two (on each side of the throat) disks, revealing a rich multi-photon ring structure due to light crossing the throat, and a strong reduction in the size of the central brightness depression region. These new rings are more numerous and far more luminous in the two-disk case than in the single-disk case, and the shadow's size reduction far more acute, making a neat distinction as compared to canonical black hole images. These results highlight the potential of high-resolution imaging in providing smoking guns for the existence of ultra-compact objects distinct from black holes via their multi-ring structure.

\end{abstract}

\maketitle

\section{Introduction}

Understanding the observational signatures of compact objects is a critical endeavor in modern strong-field astrophysics. In recent years, astrophysical observations have provided us with two new channels to peer into the nature of these objects and, at the same time, to probe the strong-field aspects of the gravitational interaction. These correspond to the LIGO/VIRGO collaboration's
detection of gravitational waves (GWs) out of binary mergers, both black holes \cite{Abbott:2016blz} and neutron stars \cite{Abbott:2021xx}, and
the Event Horizon Telescope (EHT) images of the shadows of supermassive black holes \cite{EHT:2019apjlM87,EHT:2022apjlSgrA}, observations providing strong evidence supporting both the existence of black holes and the validity of General Relativity (GR) itself. These channels belong to the newly-born field of {\it multimessenger astronomy}, namely, astronomy with several carriers: cosmic rays, neutrinos, light, gravitational waves, and shadows \cite{Addazi:2021xuf}.

These breakthroughs also pave the way for investigating the potential existence of further compact objects, either under the form of modified black holes such as hairy ones \cite{Herdeiro:2015waa} or through  alternative horizonless objects that may act as black hole mimickers \cite{Bambi:2025wjx}, via careful analysis of subtle observational features using the above tools. For instance, although the inspiral, merger, and ringdown phases of binary mergers are all well described in terms of waveform templates from (canonical, GR-based) black holes with specific masses and spins, and the optical images observed from the M87 and Milky Way (Sgr A$^*$) galaxies are also consistent with the results of General Relativistic MagnetoHydroDynamics (GRMHD) simulations of the hot, highly-magnetized accretion flow around a black hole, the existence
of other compact objects that could produce similar waveforms and shadows
cannot be definitively excluded yet. Of particular interest from this observational perspective are ultra-compact objects (UCOs), defined as those which hold surfaces of unstable bound geodesics in which GWs and shadows are naturally generated. This family includes as prominent members boson stars \cite{Liebling:2012fv} and wormholes \cite{Visser:1995cc}, among many others \cite{Mazur:2001fv,Coleman:1985ki,Mathur:2005zp,Visinelli:2021uve,Simpson:2021dyo}, to which the seemingly old-fashioned but still very effective techniques of canonical astrophysics have strongly constrained their properties \cite{Cardoso:2019rvt}.

The main aim of this work is to study the observational appearance of one such object, namely, reflection-asymmetric thin-shell, traversable wormholes. A wormhole is a solution of the gravitational field equations allowing to connect two portions of two space-times via a throat. When the wormhole is traversable, i.e., in the absence of event horizons, then a two-way transfer of energy and particles across the throat is possible. Traversable wormhole models can be built through different techniques e.g. by promoting the radial coordinate to a radial function like in black bounce space-times \cite{Simpson:2018tsi,Franzin:2021vnj}, or by using the junction conditions formalism \cite{Darmois,Israel:1966rt}. In the latter approach, two space-times are matched at a certain hypersurface (dubbed within the wormhole field as the thin-shell or throat), with the matter fields satisfying certain conditions both at the shell and across of it. Furthermore, the space-times at each side of the throat do not need to be the same (i.e. they can correspond to different solutions of the gravitational field equations). In the latter case, we have a wormhole lacking reflection-symmetry across the throat, a possibility that was introduced via a specific ad hoc construction in \cite{Wielgus:2020uqz}. It is worth pointing out that while in the framework of GR any such construction is doomed to violate the energy conditions (a trivial consequence of the singularity theorems \cite{Senovilla:2014gza}), this is not necessarily so when one switches to a modified gravitational theory because the new gravitational field equations will typically unlink the correspondence between the null (time-like) congruence condition and the null (weak) energy condition, the former being required by the singularity theorems.

This exciting possibility was recently given explicit support within the context of theories of gravity of the $f(R)$ type coupled to electromagnetic fields and formulated in the Palatini approach. The latter is a framework for gravity in which the metric and connection are regarded as independent fields, such that their corresponding field equations are obtained through independent variations of the action with respect to each of them \cite{Olmo:2011uz}. It has been shown that, for a large family of theories of gravity constructed from scalar contractions of the (symmetric part of the) Ricci tensor of a torsionless connection (examples of these theories are GR itself, $f(R)$, quadratic gravity, or Eddington-inspired Born-Infeld gravity \cite{Banados:2010ix,BeltranJimenez:2017doy}), they give rise to ghost-free, second-order field equations \cite{Afonso:2017bxr,Afonso:2018bpv,BeltranJimenez:2020sqf}, which contrast with the troubles found by their metric counterparts which require strong restrictions over the new components of the action \cite{Sotiriou:2008rp}. Many phenomenological applications of these theories have been discussed in the literature, see e.g. \cite{Enckell:2018hmo,Shaikh:2018yku,Shao:2020weq,Zeng:2025xoe}. For the sake of this work we are interested in the family of asymmetric wormholes built within this framework in Ref. \cite{Lobo:2020vqh}. This family has branches of solutions which are supported by matter sources satisfying the energy conditions everywhere and that are linearly stable. This provides a suitable, well-grounded scenario to investigate their observational signatures.

The main aim of this work is to employ the principles of geodesic integration and ray-tracing techniques to characterize the optical appearance of these asymmetric wormholes by a thin-accretion disk. In the context of generation of images, black holes generically display two features: an outer bright ring of radiation, and a central brightness depression \cite{Gralla:2019xty} (see \cite{Lupsasca:2024wkp} for a review). The luminous region is usually decomposed into several contributions (as long as the disk is optically thin), namely the disk's direct emission and a series of stacked, self-similar, highly-lensed  photon rings, characterized by the number $n$ of half-turns performed by the photons around the black hole, and which follow several scaling relations driven by a set of critical exponents entailing an exponential fall-off of successive photon rings in their widths and flux densities \cite{Gralla:2019drh,Kocherlakota:2023qgo,Kocherlakota:2024hyq}. This sequence of photon rings approaches, in the limit $n \to \infty$, a {\it critical curve} in the observer's plane image, corresponding to the projection of the photon's sphere there. While the region bounded by the critical curve is usually dubbed the black hole {\it shadow} \cite{Falcke:1999pj}, the latter only coincides with the actual brightness depression in spherical accretion models \cite{Narayan:2019}. Many studies in the literature have characterized the finer details of this picture according to the interaction between the background geometry and the accretion disk physics, see e.g. \cite{Lara:2021zth,Lima:2021las,Ozel:2021ayr,Guo:2021wid,Vincent:2020dij,Vagnozzi:2022moj,Staelens:2023jgr,Urso:2025gos}.  

The reflection-asymmetric wormholes considered here will retain these main features while introducing additional sets of higher-order photon rings not present in black hole images, and which come from light rays that travel through the other side of the wormhole throat. This multi-photon ring structure breaks the self-similarity rule of canonical photon rings and, as such, they can act as observational discriminators between asymmetric wormholes (in fact, any UCO) and canonical black holes in future upgrades of very-long baseline interferometry (VLBI). Furthermore, this sequence of new photon rings may also affect the size of the central brightness depression, which is yet another observational target of VLBI projects. And finally, the multi-photon ring structure and the reduction of shadow's size is neatly different between the case in which a single disk (on the observer's side) is present, and the two-disks case (on both sides), with the latter having significantly far more numerous and brighter rings and a drastic reduction on the shadow's size, thus significantly augmenting its chances for observational detection. The results found here are added to the pool of works in the literature \cite{Olivares:2018abq,Paul:2019trt,Qin:2020xzu,KumarWalia:2024yxn,deSa:2024dhj,Murk:2024nod,Igata:2025glk,Rosa:2022tfv,Rosa:2022toh,Olmo:2023lil,Rosa:2023qcv,Combi:2024ehi,He:2025qmq,Zhao:2025yhy} by which UCOs display new features in photon rings and shadows acting as observational discriminators compared to black holes. 

This work is organized as follows. In Sec. \ref{C:II} we recap the theoretical framework providing support to these asymmetric wormholes as given by Palatini $f(R)$ gravity, and set the specific choices for the parameters defining the two joined space-times. In Sec. \ref{C:III} we provide a discussion on the origin of the multi-ring structure by individual and bunches of light rays. In \ref{C:IV} we introduce our disk modeling and assumptions, and discuss the multi-ring structure when there is a single disk located on the observer's side, and when a second disk is located on the other side of the wormhole. In Sec. \ref{C:V} we provide our conclusions and further thoughts on the chances to the observational detectability of these wormholes.

\section{Theoretical framework}\label{C:II}

\subsection{Gravity and matter models}

Our theoretical framework is given by the action
\begin{equation} \label{eq:action}
    \mathcal{S}=\frac{1}{2\kappa^2} \int d^4x \sqrt{-g} f(R) + \mathcal{S}_m(g_{\mu\nu},\psi_m),
\end{equation}
with the following definitions and conventions: $\kappa^2=8\pi G/c^4$ is Newton's constant (hereafter we shall work in units $G=c=1$), $g$ is the determinant of the space-time metric $g_{\mu\nu}$, $f(R)$ is a given function of the curvature scalar $R=g^{\mu\nu}R_{\mu\nu}(\Gamma)$, where the Ricci tensor $R_{\mu\nu}(\Gamma) \equiv {R^\alpha}_{\mu\alpha\nu}(\Gamma)$ is built from the Riemann tensor
\begin{equation}
    R^\alpha_{\ \beta \mu \nu} = \partial_\mu \Gamma^\alpha_{\nu \beta} - \partial_\nu \Gamma^\alpha_{\mu \beta} + \Gamma^\alpha_{\mu \lambda} \Gamma^\lambda_{\nu \beta} - \Gamma^\alpha_{\nu \lambda} \Gamma^\lambda_{\mu \beta},
\end{equation}
and is solely dependent on the (torsionless) affine connection, $\Gamma \equiv \Gamma_{\mu\nu}^{\lambda}$. As for the matter sector, $\mathcal{S}_m=\int d^4x \sqrt{-g} \mathcal{L}(g_{\mu\nu},\psi_m)$, it couples the set of matter fields $\psi_m$ only to the metric and not to the connection, in order to fulfill the equivalence principle (i.e. the equality of all frames for free-fall motion).

The action in Eq. (\ref{eq:action}) is a simple extension of GR via additional corrections in the curvature scalar, suitably suppressed by some inverse length-squared scale. It is at this point that the Palatini approach to gravity kicks in, as provided by the fact that the affine connection $\Gamma$ is taken as independent of the metric tensor $g_{\mu\nu}$. In this approach, therefore, the corresponding field equations are obtained by independent variations of Eq. (\ref{eq:action}) with respect to each independent quantity, which provides the following two sets \cite{Olmo:2011uz}:
\begin{eqnarray}
f_R R_{\mu\nu} - \frac{1}{2}g_{\mu\nu}R&=& \kappa^2 T_{\mu\nu} \label{eq:metric}, \\
\nabla_{\mu}^{\Gamma}(\sqrt{-g} f_R g^{\alpha\beta})&=&0 \label{eq:connection},
\end{eqnarray}
where $\nabla_{\mu}^{\Gamma}$ is the covariant derivative defined in terms of the independent connection, we have defined $f_R \equiv df/dR$, and $T_{\mu\nu} =-\tfrac{2}{\sqrt{-g}} \tfrac{\delta \mathcal{S}_m}{\delta g^{\mu\nu}}$ is the energy-momentum tensor of matter. It is worth stressing that, for the Einstein-Hilbert action of the GR case, $f(R)=R$, Eq. (\ref{eq:connection}) for the connection reduces to the usual metric-connection compatibility equation, telling us that the connection is given by the Christoffel symbols of the metric. Furthermore, Eqs. (\ref{eq:metric}) for the metric reduce to the usual Einstein field equations and, consequently, the Palatini formulation of GR is equivalent to the usual metric one of setting the Levi-Civita connection {\it a priori}\footnote{Modulo a projective mode which has no consequences neither for the solutions of the theory nor for the motion of particles \cite{Bejarano:2019zco}, but may have an impact for the sake of boundary terms and quantities derived from them \cite{BeltranJimenez:2019esp}.}. However, this is generally not true for actions beyond the Einstein-Hilbert one and, in particular, for the $f(R)$ theories supporting the wormhole geometries considered in this work. 

\subsection{Junction conditions and thin-shells}

At this stage one could solve Eqs. (\ref{eq:metric})  and (\ref{eq:connection}) in the usual way, that is, by setting a matter Lagrangian density and a given symmetry for the line element, and solving the combination of connection and metric equations. Furthermore, one can use the opportunities provided by the {\it junction conditions} method to build further solutions of interest from previous ones. In this method, introduced by Darmois \cite{Darmois} and Israel \cite{Israel:1966rt}, two space-times (as given by their line elements), which are assumed to be solutions of a given
theory, are joined (matched) at a certain hypersurface. The enlarged solution is intended to have interesting properties of its own and, in particular, this formalism is suitable to build configurations in which there is a sharp transition between two (or more) regions, like in stars inside/outside their surfaces, in gravastars or, for the case of interest of this work, in traversable wormholes.

The junction conditions formalism for Palatini $f(R)$ theories was developed in \cite{Olmo:2020fri} and is summarized in Appendix \ref{A:JC}.  It is based on the use of tensorial distributions to find the allowed discontinuities of the geometrical quantities across the matching hypersurface without introducing divergences in the field equations induced by derivatives of the tensorial distributions. At the same time, it allows to find the features of the required matter fields (if any) at the matching hypersurface from the energy conservation equations. The application of this formalism to Palatini $f(R)$ gravity introduces some significant novelties compared to its metric counterparts \cite{Senovilla:2013vra} (and to GR itself), opening new avenues to play with the corresponding solutions and their phenomenology. In the case of spherically symmetric thin-shell wormholes, this formalism allows to fix the equation of state on the shell in order to build scenarios of interest e.g. that of traceless energy-momentum tensors, such as in the case of solutions supported by an electric Maxwell field.

\subsection{Asymmetric thin-shell wormholes and model parameters} \label{sec:asym}

We consider two surgically joined Reissner-Nordstr\"om space-times $\mathcal{M}_+$ and $\mathcal{M}_-$ of the form
\begin{eqnarray}
ds_{\pm}^2&=&-A_{\pm}(r)dt^2 + \frac{dr^2}{A_{\pm}(r)}+r^2 d\Omega^2, \\
A_{\pm}(r)&=&1-\frac{2M_{\pm}}{r} + \frac{Q_{\pm}^2}{r^2},
\end{eqnarray}
where $d\Omega^2=d\theta^2+\sin^2 \theta d\phi^2$ is the line element in the two-spheres, while $M_{\pm}$ and $Q_{\pm}$ label the $\mathcal{M}_{\pm}$ space-times masses and charges, respectively, and which are allowed to be different from each other. These geometries are matched on a spherical surface of radius $r_0$ to form a single manifold $\mathcal{M}=\mathcal{M}_+ \cup \mathcal{M}_-$, dubbed {\it asymmetric wormhole} since the geometry is different on each side of the throat.

This construction is possible within the Palatini formalism, as shown in \cite{Lobo:2020vqh}. Furthermore, this is true for any polynomial $f(R)$ model, like the quadratic one, $f(R)=R+\alpha R^2$ (with constant $\alpha$). This is so because, in electro-vacuum solutions of Maxwell type, due to the vanishing of the trace of the energy-momentum tensor $T_{\mu\nu}$, Palatini $f(R)$ solutions reduce to those of GR, while it is the new dynamics of the gravity and matter fields at the shell created by this approach that are responsible for generating these solutions, as described in \cite{Guerrero:2021pxt}. Furthermore, in this model the matter sources at the shell are given by an energy density and pressure which take the form
\begin{equation} \label{eq:eos}
\sigma=\frac{C}{r_0^3} \quad ; \quad P=\frac{\sigma}{2} \ , 
\end{equation}
respectively, where $C$ is a constant which is determined by one of the components of the junction condition (\ref{JC4}). It turns out that there is a tiny gap in the parameter space (as given by the masses and charges on each side) where the corresponding traversable wormholes are linearly stable and supported by a positive energy density at the shell. This way, these traversable thin-shell wormholes fulfill all the energy conditions both outside the shell (since they are supported by usual Maxwell fields) and at the shell due to this positive energy density there.

In our work we assume, without any loss of generality, the observer to be located on the $\mathcal{M}_+$ side. In order to cast the problem in suitable form, we introduce the following definitions:

\begin{itemize}
\item $\xi=M_+/M_-$: The mass-to-mass ratio between  the space-times at each side of the wormhole.
\item $\eta=Q_+^2/Q_-^2$: The quadratic charge-to-charge ratio between each side. 
\item $y=Q_-^2/M_-^2$: The quadratic charge-to-mass ratio on the $\mathcal{M}_{-}$ side.
\item $x_0=r_0/M_{-}$: The radius of the shell (in units of the $M_{-}$ mass).
\end{itemize}
Furthermore, from the condition imposed by the junction conditions formalism regarding the continuity of the metric across the shell, one finds such quantities to be weaved together via the equation
\begin{equation}
\xi=1-\frac{y}{2x_0}(1-\eta).
\end{equation}
With these definitions, the horizons on each side of the wormhole (if any) are given by (for convenience, we write here all radial locations in terms of the dimensionless coordinate $x \equiv r/M_{-}$)
\begin{eqnarray}
x_h^{-}&=&1+\sqrt{1-y},  \\
x_h^{+}&=&\xi(1+\sqrt{1-\eta y/\xi^2}).  
\end{eqnarray}
We also need expressions for the location of the photon sphere (if any), namely, the surfaces of unstable bound photon orbits, given the key role they play in the generation of images, as described in Appendix \ref{A:PO}. For these configurations, the photon spheres are located at 
\begin{eqnarray}
x_{ps}^{-}&=& \frac{3+\sqrt{9-8y}}{2},  \\
x_{ps}^{+}&=&\xi\left(\frac{3+\sqrt{9-8\eta y/\xi^2}}{2} \right).
\end{eqnarray}

The tiny gap in the space-parameter for linearly-stable, positive-energy density thin-shell wormholes is described in Fig. 2 of Ref. \cite{Guerrero:2021pxt}. For the sake of this work we take the following values for the model parameters 
\begin{eqnarray}
\eta&=&2, \\
x_0&=&1.5, \\
y&=&0.95, \\
\xi & \approx & 1.316 \label{eq:xi}.
\end{eqnarray}
which fall within that gap. These parameters actually correspond to the values taken in the example of Ref. \cite{Guerrero:2021pxt}. For this choice, one has $x_h^- \approx 1.316$, while no solution exists for $x_h^+$. This means that the would-be horizon (on the $\mathcal{M}_{-}$ side) is below the wormhole's throat location, $x_h^+<x_0$, while the would-be naked singularity on the  $\mathcal{M}_{-}$ side is effectively removed in the matching process. Therefore, we have a traversable thin-shell wormhole scenario, allowing fluxes of energy and particles throughout the throat on both directions. Furthermore, each side of the wormhole is compact enough to allow for photon spheres. Their locations and corresponding critical impact parameters are given by (hereafter we work in units of $M_-=1$ and $M_+ = \xi$, recall the above definitions and Eq.(\ref{eq:xi}))
\begin{eqnarray}
   x_{ps}^{-}&=&2.09161 \quad ; \quad b_c^- \approx 4.095, \\
   x_{ps}^{+}&=&2.29221 \quad ; \quad b_c^+ \approx 4.969.
\end{eqnarray}
For comparison, in a Schwarzschild space-time one has $x_{ps}=3M, b_{c}=3\sqrt{3}M$, expressed in units of the asymptotic space-time mass $M$ of the black hole. We shall next employ the above picks in order to carry out the ray-tracing procedure prior to the generation of images in our work.

\section{Ray-tracing procedure} \label{C:III}

\begin{figure}[t!]
\includegraphics[width=8.0cm,height=5.5cm]{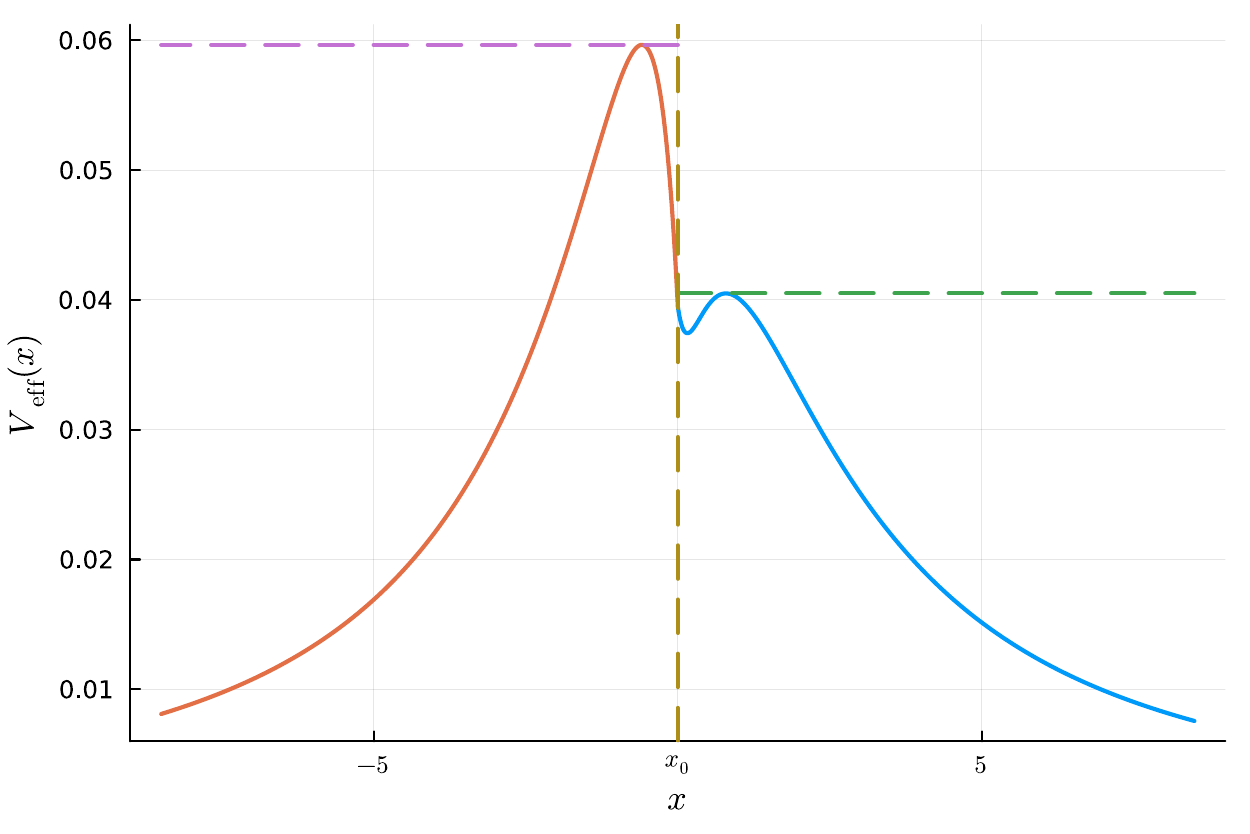}
\caption{Effective potential $V_{eff}(x)$ as given by the union of the two potentials of Eq.(\ref{eq:effpot}) for the asymmetric wormhole according to the parameters set in Sec. \ref{sec:asym}. Both parts of the potential meet at the throat (yellow dashed line). The potential has two maxima: one in $\mathcal{M}_-$ (magenta dashed line), and one in $\mathcal{M}_+$ (green dashed line). We consider observers located on the right-hand side of this figure, which are the only ones capable of seeing the multi-photon ring structure.}
\label{fig:effective_pot}
\end{figure}

The above setting is thus a traversable thin-shell wormhole with different photon spheres at each side of the throat. The corresponding effective potentials are written as (recall Appendix \ref{A:PO} for basic definitions and concepts)
\begin{equation} \label{eq:effpot}
V_{\pm}=\frac{A_{\pm}}{r^2},
\end{equation} 
and the full potential resulting from joining these two pieces is depicted in Fig. \ref{fig:effective_pot}, where we observe the presence of two maxima, one on $\mathcal{M}_{+}$ and another on $\mathcal{M}_{-}$, each of them corresponding to their photon spheres though with different locations and heights. When a photon sphere is present in the gravitational field of an UCO, photons issued from a luminous source (e.g. a star) and whose impact parameter nears the critical one, will experience large gravitational deflection when approaching the photon sphere \cite{Perlick:2021aok}. As a consequence, photons can execute several half-turns around the UCO, the latter defined via the equation
\begin{equation} \label{eq:ndef}
\phi(r) = \frac{\pi}{2} + n \pi \quad \text{with} \quad n \in \mathbb{N},
\end{equation}
and corresponding to successive images of the front and back of the disk. The first image, $n=0$ corresponds to the disk's direct emission, while $n=1,2,\ldots$ are collectively referred to as the {\it photon ring}. The photon ring creates, on the observer's screen, a local, narrow boost of luminosity. Furthermore, in the optically-thin limit (i.e. absence of significant absorption), and as long as there are ``gaps" in the emission region \cite{Vincent:2022fwj}, the photon ring is decomposed into an infinite set of self-similar rings indexed by the number $n$ of half-turns. Such a sequence approaches, in the limit $n \to \infty$, a {\it critical curve} corresponding to the projection, on the observer's screen, of the photon sphere\footnote{In the black hole case, variations in the photon's shape with the surrounding emission \cite{Gralla:2020srx,Paugnat:2022qzy}, and exponential-like decays driven by Lyapunov exponents \cite{Wielgus:2021peu}  could be used, via VLBI observations, to infer the geometrical configuration generating such photon rings via its interferometric signatures \cite{Johnson:2019ljv,Cardenas-Avendano:2023dzo}.}. For finite $n$, the corresponding ring may appear overlapped with the direct emission or separately from it, depending on the features of the emission model (more specifically, on its effective region of emission).

To trace the light rays of our thin-shell wormhole with two photon spheres bearing the ingredients above, we assume an observer located on the $\mathcal{M}_{+}$ side of the wormhole. This is because it is the only observer capable of seeing new photon rings, due to its capability to optically map the intermediate region between both maxima. To track this, we perform a backwards ray-tracing procedure, in which every light ray arriving at the observer's screen (assumed to be located at asymptotic infinity) is traced backwards to its emitting source location. This implies integrating the geodesic equation, suitably written as
\begin{equation}\label{eq: geodesic}
\frac{d\phi_\pm}{dr}= \mp \frac{b}{r^2} \frac{1}{\sqrt{1-b^2V_{\pm}(r)}},
\end{equation}
where the signs $\pm$ refer to outgoing (ingoing) trajectories in the $\mathcal{M}_+$ ($\mathcal{M}_{-}$) side, while $\phi_{\pm}$ refer to the deflection angle turned on each $\mathcal{M}_{\pm}$ side. 

This way we find three possible scenarios for the result of the ray-tracing integration procedure, in which we proceed as follows:

\begin{enumerate}

\item Photons with $b>b_c^+$ only see the effective potential $V_{+}(r)$ in $\mathcal{M}_{+}$ and are integrated with the $(-)$ sign until arriving to the turning point $r_{tp}$ satisfying $V_{+}(r_{tp})=1/b^2$, at which the integration flips sign to $(+)$ and continues back to asymptotic infinity in $\mathcal{M}_{+}$.

\item Photons with $b<b_c^{-}$ see first the effective potential $V_{+}(r)$ in the region $\mathcal{M}_{+}$ and are integrated with the $(-)$ sign until crossing the wormhole throat $r_0$ to $\mathcal{M}_{-}$. From there, the integration is carried out with the ($+$) sign in the effective potential $V_{-}(r)$ but a turning point is never found: instead these trajectories exit the photon sphere $r_{ps}^-$ and keep their trip towards asymptotic infinity in the $\mathcal{M}_-$ geometry.  

\item Photons with $b_{c}^{-}<b<b_{c}^{+}$ see first the effective potential $V_{+}(r)$ in the region $\mathcal{M}_{+}$ and are integrated with the $(-)$ sign there, traverse the photon sphere $r_{ps}^+$ and eventually cross the throat's location $r=r_0$. In the region $\mathcal{M}_{-}$ these are outgoing trajectories  and see the effective potential $V_{-}(r)$ so the integration is carried out with the ($+$) sign. Eventually the photon hits a turning point $r_{tp}$ satisfying $V_{-}(r_{tp})=1/b^2$, at which the integration flips sign to $(-)$ but it is still carried out using $V_{-}(r)$ since it remains on $\mathcal{M}_{-}$. Later, the photon crosses back the throat and enters the original region $\mathcal{M}_{+}$, in which it sees the potential $V_{+}(r)$ again and, being an outgoing trajectory there, the integration is carried out with the ($+$) sign until reaching asymptotic infinity.

\end{enumerate}

\begin{figure*}[t!]
\includegraphics[width=8.0cm,height=5.5cm]{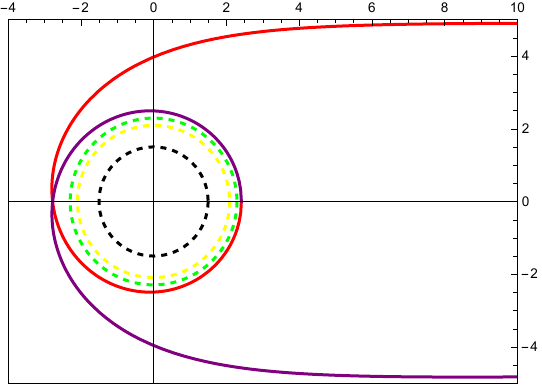}
\includegraphics[width=8.0cm,height=5.5cm]{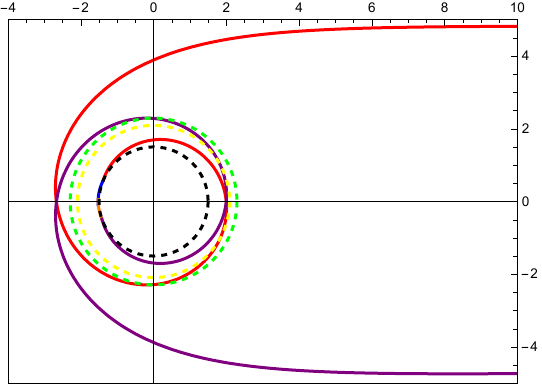}
\includegraphics[width=8.0cm,height=5.5cm]{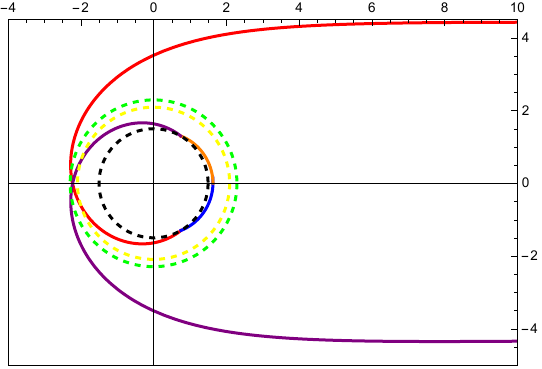}
\includegraphics[width=8.0cm,height=5.5cm]{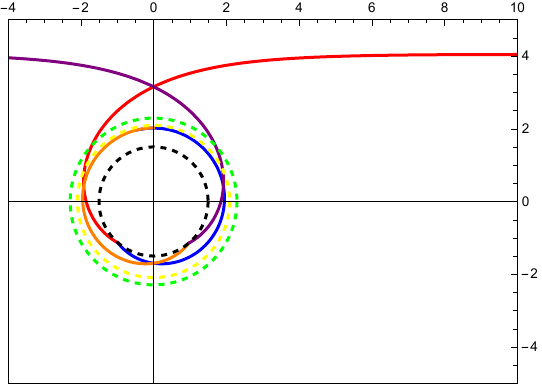}
\caption{Trajectories of individual light rays in  asymmetric traversable  thin-shell wormholes for four picks of impact parameters: $b=b_c^+ + 0.0083$ (top left), $b=b_c^+ -0.0775$ (top right), $b=b_c^-+0.4$ (bottom left), and $b=b_c^-+0.005$ (bottom right), corresponding to the model parameters and units discussed in  Sec. \ref{sec:asym}. Red  curves correspond to ingoing trajectories on $\mathcal{M}_+$; blue curves to outgoing trajectories in $\mathcal{M}_-$;  orange curves to ingoing trajectories in $\mathcal{M}_-$; and purple curves to outgoing trajectories in $\mathcal{M}_+$. The dashed black circle corresponds to the location of the wormhole throat, while the dashed green (yellow) circle corresponds to the location of the photon sphere on $\mathcal{M}_+$($\mathcal{M}_-$). Note that, the closer the impact parameter to $b_c^+$ ($b_c^-$) the larger the number of half-orbits on $\mathcal{M}_+$($\mathcal{M}_-$).}
\label{fig:metric}
\end{figure*}

\begin{figure}[t!]
    \centering    \includegraphics[width=1\linewidth]{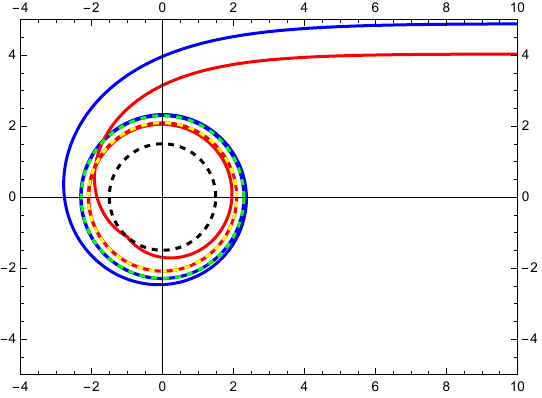}
    \caption{Ray-tracing of two null geodesics with $b\approx b_c^\pm$ (blue and red), each approaching the corresponding photon sphere on the $\mathcal{M}_+$ and $\mathcal{M}_-$, respectively. On their path, each photon performs an arbitrary large number of half-turns.}
\label{fig:critical_ph}
\end{figure}

\begin{figure*}[t!]
\centering
\includegraphics[width=6.8cm,height=4.8cm]{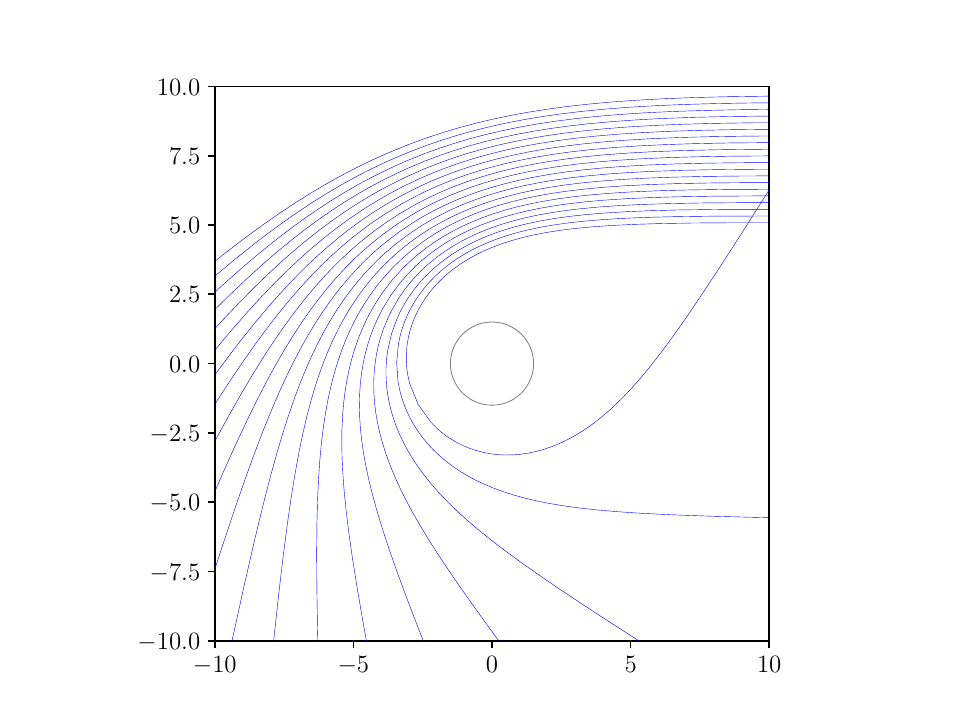}%
\hspace{-1.3cm}%
\includegraphics[width=6.8cm,height=4.8cm]{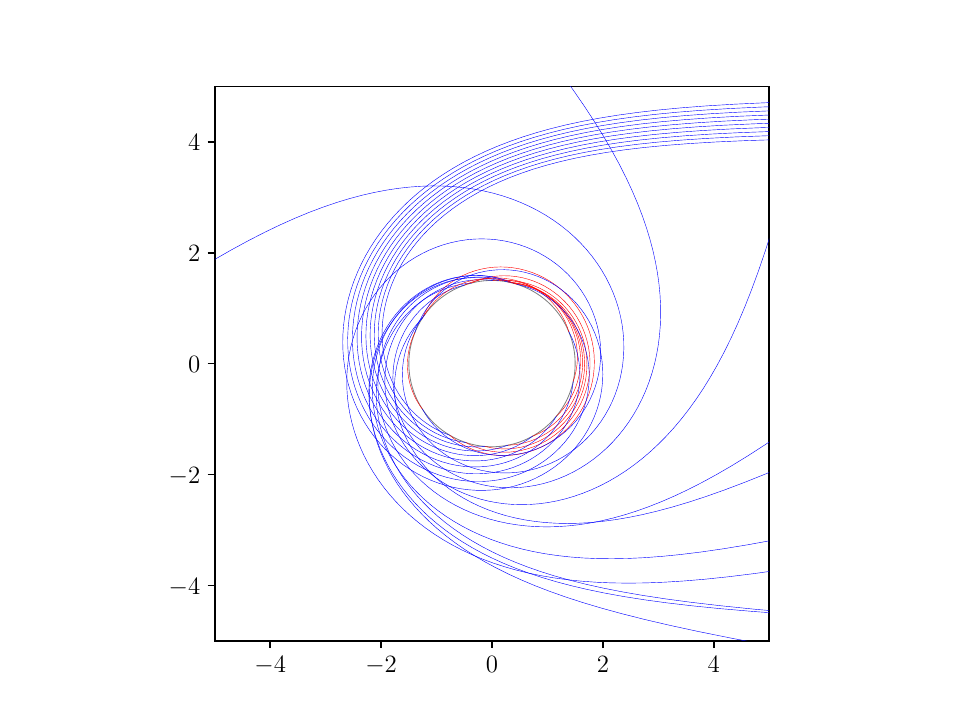}%
\hspace{-1.3cm}%
\includegraphics[width=6.8cm,height=4.8cm]{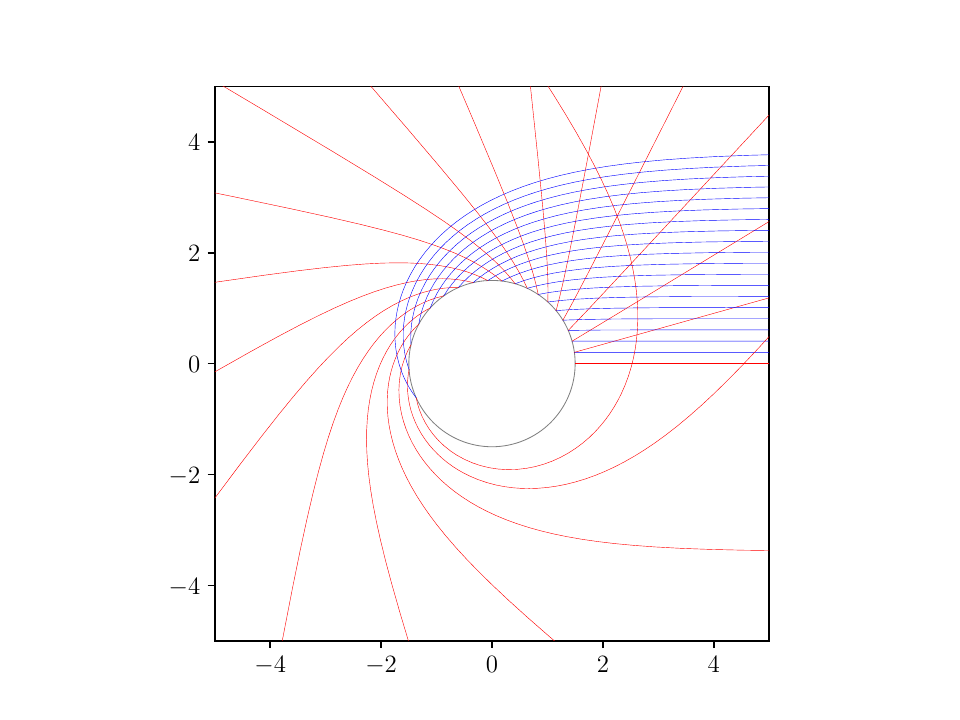}%
\caption{Ray-tracing for $b_c^+\leq b \leq 10$ (left), $b_c^-\leq b \leq b_c^+$ (middle), $0\leq b \leq b_c^-$ (right), representing the three possible behaviours of light rays according to the regions of the full manifold $\mathcal{M}$ they explore. Blue (red) curves correspond to trajectories on $\mathcal{M}_+$ ($\mathcal{M}_-$), while the wormhole throat corresponds to the central circle.}
\label{fig:ray_tracing}
\end{figure*}

The first scenario corresponds to the usual black hole one. These photons contribute to the direct image of the source and, in addition, provide a series of photon rings from highly-bent trajectories nearing the photon sphere, $b \gtrsim b_c^{+}$, and which appear in the observer's screen as self-similar images of the source.

The second scenario above corresponds to those light rays that cross the wormhole throat but never come back to the region $\mathcal{M}_+$ they were issued from, going instead to the asymptotic infinity of the region $\mathcal{M}_-$. This is the equivalent, in the black hole case, of those photons that intersect the event horizon and give rise to the black hole shadow.

The third scenario is a genuine contribution of the thin-shell wormhole scenario. These are the light rays whose impact parameter allows them to traverse the wormhole throat towards $\mathcal{M}_{-}$, find a turning point in the effective potential $V_{-}(r)$ and cross back the throat towards its original space-time $\mathcal{M}_{+}$. Furthermore, light rays issued with an impact parameter $b \lesssim b_c^+ $ ($b \gtrsim b_c^{-}$) correspond to highly-bent trajectories in the internal (external) part of the $\mathcal{M}_{+}$ ($\mathcal{M}_{-}$) photon sphere $x=x_{ps}^{+}$ ($x=x_{ps}^{-}$), which we recall are associated to the two maxima displayed in Fig. \ref{fig:effective_pot}.

In Fig. \ref{fig:metric} we depict the result of the ray-tracing procedure for four individual rays with carefully selected impact parameters to highlight the above discussion: one slightly above the $\mathcal{M}_{+}$ critical impact parameter, $b \gtrsim b_c^{+}$ (top left), one slightly below of it, $b \lesssim b_c^{+}$ (top right), one in the region between the critical impact parameters of the $\mathcal{M}_{\pm}$ regions,  $b_c^{-} < b < b_c^{+}$ (bottom left), and another one slightly above the $\mathcal{M}_{-}$ critical impact parameter, $b \gtrsim b_c^{-}$ (bottom right). In the first case there are only ingoing (red) and outgoing (purple) trajectories in $\mathcal{M}_{+}$, typical of a black hole space-time, while in the other three cases we find ingoing and outgoing trajectories in $\mathcal{M}_{+}$ but also outgoing (blue) and ingoing (orange) trajectories in $\mathcal{M}_{-}$. To further highlight the role played by the two photon spheres, in Fig. \ref{fig:critical_ph} we depict the ray-tracing for two light rays whose impact parameters are very near the critical ones, $b \approx b_c^{\pm}$. In both cases the light ray approaches the corresponding photon sphere (on $\mathcal{M}_{\pm}$, respectively) performing an arbitrarily large number of half-turns.

In Fig. \ref{fig:ray_tracing} we depict the result of the ray-tracing procedure for a bunch of light rays whose impact parameters cover three different ranges: $b_c^+ < b$ (left), $b_c^{-}<b<b_c^{+}$ (middle), and  $0<b< b_c^{-}$ (right), traversing through the region $\mathcal{M}_{+}$ (blue) or $\mathcal{M}_{-}$ (red). In the first case we have the usual contributions to the image of a black hole including the highly-bent trajectories that contribute to the photon ring images on the $\mathcal{M}_{+}$ side. In the second case, we have trajectories that travel through the $\mathcal{M}_{-}$ side between the two maxima of the photon sphere before coming back to its original $\mathcal{M}_{+}$ side, contributing to further images and new photon rings. In the third case we find trajectories that go through the throat to the region $\mathcal{M}_{-}$ side but are not capable of turning back to $\mathcal{M}_{+}$.

Following this behavior of individual and bunches of light rays, we find the expected following novelties as compared to a typical black hole scenario:

\begin{itemize}

\item There will be new photon rings associated to photons that pass close to any of the photon spheres $x_{ph}^{\pm}$ within the range $b_c^- < b < b_c^+$. In the case of $b \lesssim b_c^{+}$ such photon rings can come from ingoing light rays on its trip towards the wormhole throat, or from outgoing rays from the throat towards asymptotic infinity in $\mathcal{M}_+$, while for $b \gtrsim b_{c}^-$ they are originated from ingoing/outgoing rays near the photon sphere of $\mathcal{M}_-$.

\item The typical scaling relations of critical exponents delivering an exponential decay of photon rings in their width and luminosities will be overruled by these additional photon rings, given the far more complex interplay between ingoing/outgoing trajectories.

\item A central brightness depression (i.e., the actual shadow) will still be generically present. It will be associated with those light rays that cross the throat of the wormhole and are not capable of turning back to $\mathcal{M}_+$, whose size should thus be associated to the impact parameter $b_c^-$. However, this size could be further decreased if we consider a second source of luminosity originated on the $\mathcal{M}_{-}$ side, and emitting towards the throat in order to reach the region $\mathcal{M}_{+}$.
    
\end{itemize}

The analysis above on individual and bunches of light rays provides useful information on the expected observational appearance of asymmetric traversable thin-shell wormholes, but does not take into account the fact that, in practical terms, the main source of illumination of a compact object is provided by its accretion disk, and therefore we tackle the corresponding optical appearances in the next section.

\section{Multi-photon ring structure}

\label{C:IV}

\subsection{Thin accretion disks}

The characterization of a disk orbiting an UCO comes from several parameters, including (here $\nu$ denotes the photon's frequency) the absorptivity $\alpha_{\nu}$, intensity $I_{\nu}$, and emissivity $j_{\nu}$, waved together via the radiative transfer equation \cite{Gold:2020iql}. For the sake of generation of images in our asymmetric  traversable thin-shell wormhole scenario, we shall make use of a simplified treatment of an accretion disk previously (and widely) employed in the literature and whose assumptions we summarize here:

\begin{itemize}

\item Infinitely-vanishing thickness (i.e. geometrically thin). This choice is motivated on the grounds that, according to current understanding of the formation of black hole images, the features of both the photon rings and the central brightness depression in actual disks with gaps in their emission region are more qualitatively similar to the thin-disk model than to the spherical one \cite{Vincent:2022fwj}, and it can be furthermore corrected via additional fudge factors to improve the agreement with the results of time-averaged images from GRMHD simulations \cite{Chael:2021rjo,Paugnat:2022qzy}. This assumption significantly simplifies the generation of images and reduces computational times.

\item Zero absorption (i.e. optically thin). This assumption is needed in order to allow light trajectories to turn several half-turns around the wormhole and to create sets of photon rings on the observer's plane image, otherwise only the disk's direct emission would contribute to the image. While in typical black holes images it is enough to consider up to $n=2$ trajectories given the exponential fall-off of the luminosity of successive images \cite{Kocherlakota:2023qgo}, asymmetric wormholes break that rule and we shall need to incorporate additional light trajectories which contribute non-negligibly to the image.

\item Monochromatic intensity in the disk's frame, that is, $I_{\nu}^e=I(r)$, described by a suitable adaptation of Johnson's Standard Unbound distribution 
\begin{equation}
    I(r:\gamma,\mu,\sigma) = \frac{e^{-\frac{1}{2}[\gamma + \arcsin{(\frac{r-\mu}{\sigma})}]^2}}{\sqrt{(r-\mu)^2+\sigma^2}},
\end{equation}
previously employed in the literature to reproduce time-averaged images of specific GRMHD simulations of the accretion flow around M87 and Sgr A$^*$, see e.g. \cite{Johnson:2019ljv,Paugnat:2022qzy,Cardenas-Avendano:2023dzo}. We shall use the picks for the parameters characterizing this distribution reported in Table \ref{tbl :emission},
chosen according to different emission regions and expectations for the corresponding images: GLM1 and GLM2 peak near the event horizon while GLM3 does it near the innermost stable circular orbit. These models are a restriction of those considered by Gralla, Lupsasca and Marrone (hence their name) in \cite{Gralla:2020srx}.

\begin{table}[h!]
    \centering
    \begin{tabular}{|c|c|c|c|}
        \hline
        \textbf{Source} & $\boldsymbol{\mu}$ & $\boldsymbol{\sigma}$ & $\boldsymbol{\gamma}$ \\
        \hline
        GLM1 & $x_0$ & $1.5$ & $-1.5$ \\
        GLM2 & $x_0$ & $0.5$ & $0$ \\
        GLM3 & $17x_0/6$ & $0.25$ & $-2$ \\
        \hline
    \end{tabular}
    \caption{Parameters $\mu$, $\sigma$, and $\gamma$ for the three GLM models employed in this work.}
    \label{tbl :emission}
\end{table}

\end{itemize}

Under these conditions, the observed intensity in the observer's frame can be computed as (see e.g. \cite{Olmo:2025ctf})
\begin{equation}
    I_o(r)= \sum_{i=0}^{n} \xi_i A^2(r) I(r)_{ \vert r=r_n(b)}.
\end{equation}
In this expression the summation symbol indicates the number of half-turns and it ranges from zero (the disk's direct emission) to a number $n$ of photon rings that non-negligibly contribute to the image, the function $A^2(r)$ implements the contribution of the gravitational redshift of photons, the set of numbers $\xi_i$ (fudge factors) are employed to correct the relative luminosity between the photon rings and the direct emission from the thin-disk model to the thick one, and these functions are all evaluated at $r_n(b)$, which is the {\it transfer function} marking the location on the disk a given photon of impact parameter $b$ hits (from one to multiple times due to the optically-thin assumption, reason why there are $n$ such functions). For simplicity we set $\xi_i=1, \forall i$, and we take values of the half-number of turns up to $n=7$ as a compromise between the non-negligible contributions (particularly for the two-disks case, as shown below) to the image and the need for reasonable computational times.

\subsection{Numerical setup}

Our next step is to describe our numerical setup, starting from the physical scenario in which there is a single disk on the observer's side (the case with disks at both sides will be discussed later). This is done by integrating Eq.\eqref{eq: geodesic}, considering that besides photons emitted outwards from the disk and that can reach the observer after $n$ half-turns, there are also trajectories of photons that are emitted inwards and that can traverse the throat, explore a piece of the region $\mathcal{M}_-$, turn back and come to the disk before reaching the observer in $\mathcal{M}_+$, as described in the ray-tracing procedure of Sec. \ref{C:III}. For the sake of comparison of asymmetric wormhole images with canonical black hole ones, we shall use as a benchmark for the latter a canonical charged Reissner-Nordstr\"om black hole with mass $M=1$ and charge $Q=0.5$. This choice is not intended to provide a direct quantitative comparison of observables for each case but instead a qualitative description of the differences between canonical black hole images and asymmetric wormhole ones that may act as observational discriminators.

\begin{figure*}[t!]
    \centering
    \begin{tabular}{ccc}
\includegraphics[width=5.8cm,height=4.8cm]{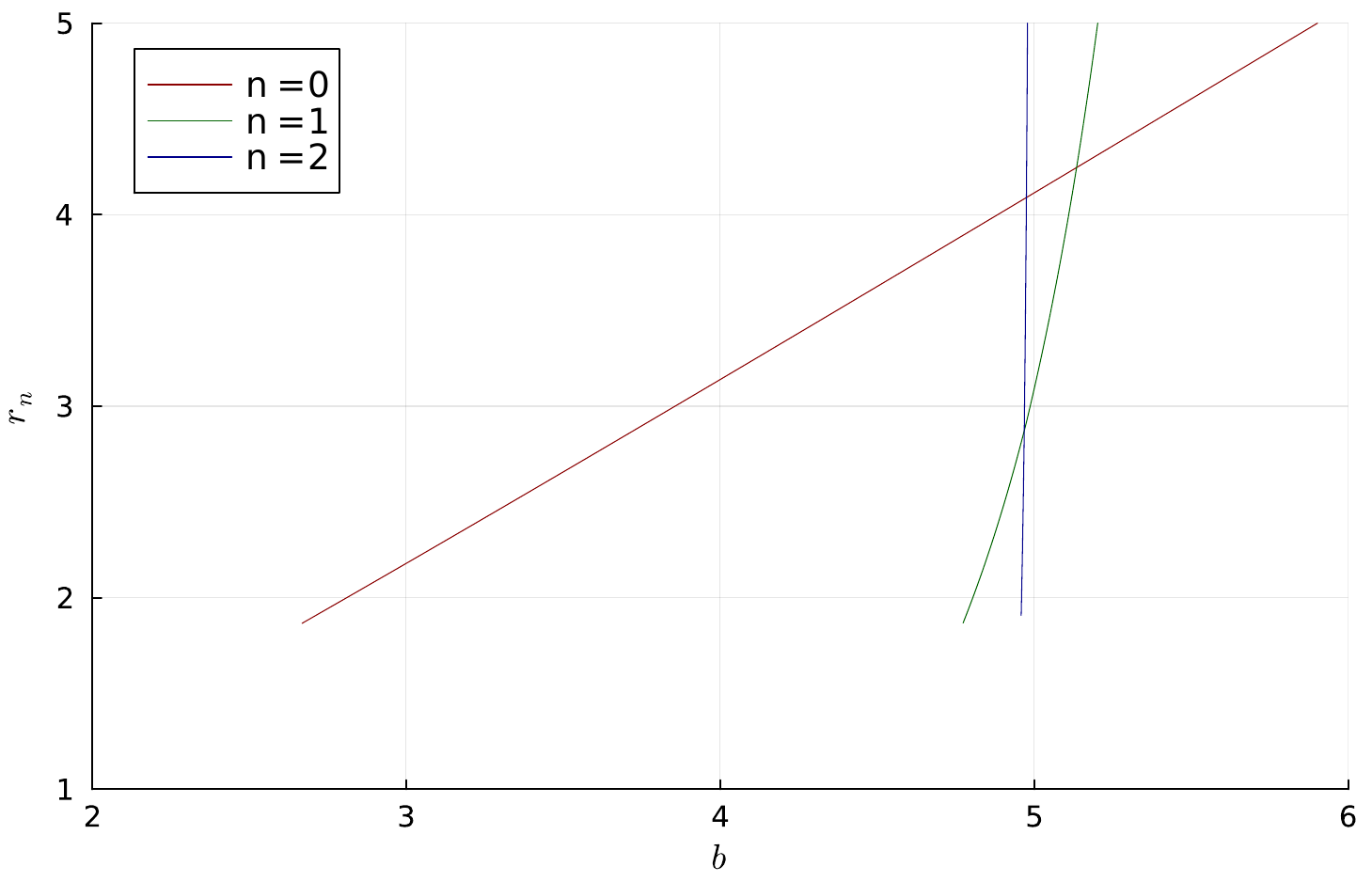} &
\includegraphics[width=5.8cm,height=4.8cm]{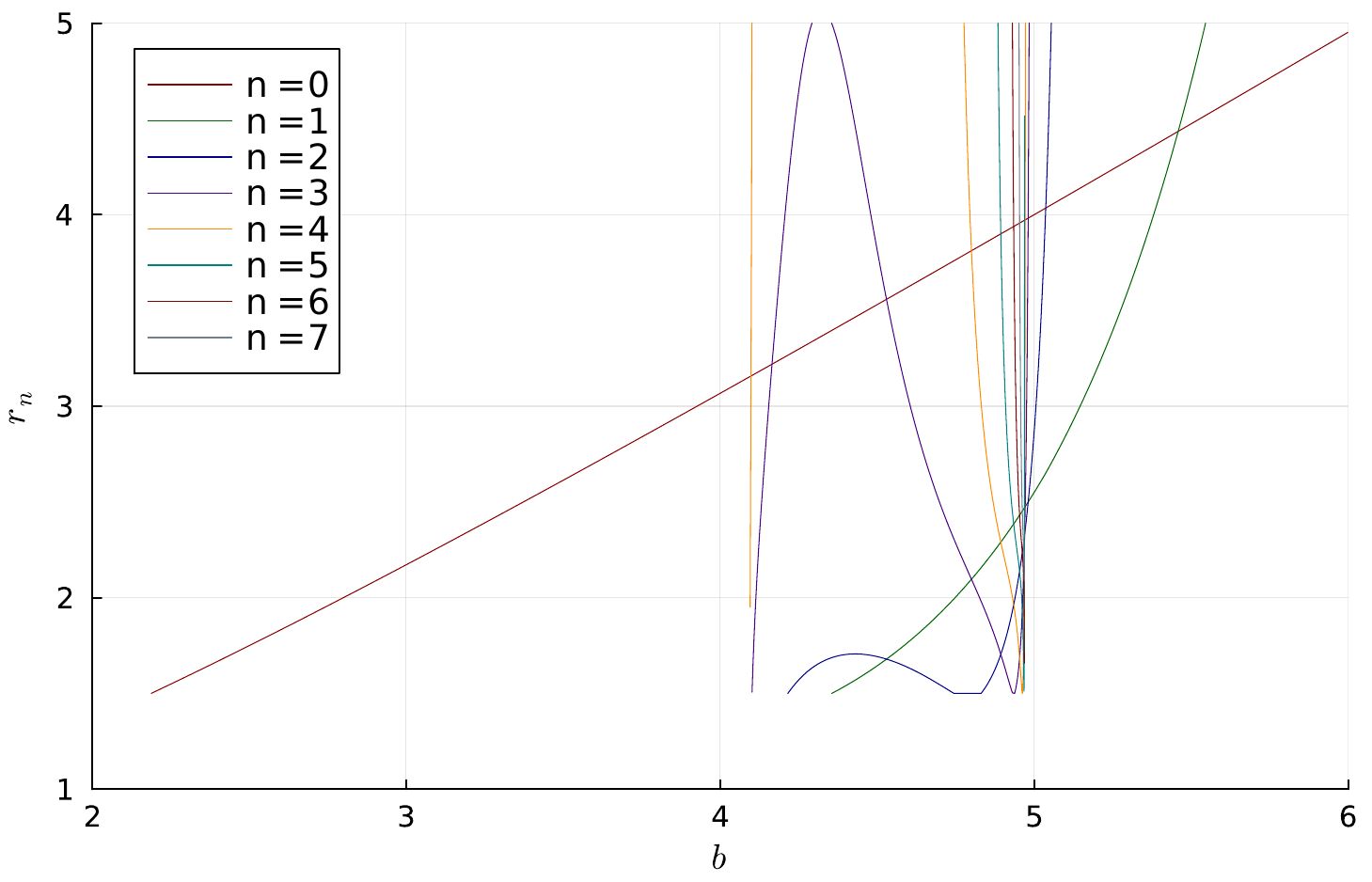} &
\includegraphics[width=5.8cm,height=4.8cm]{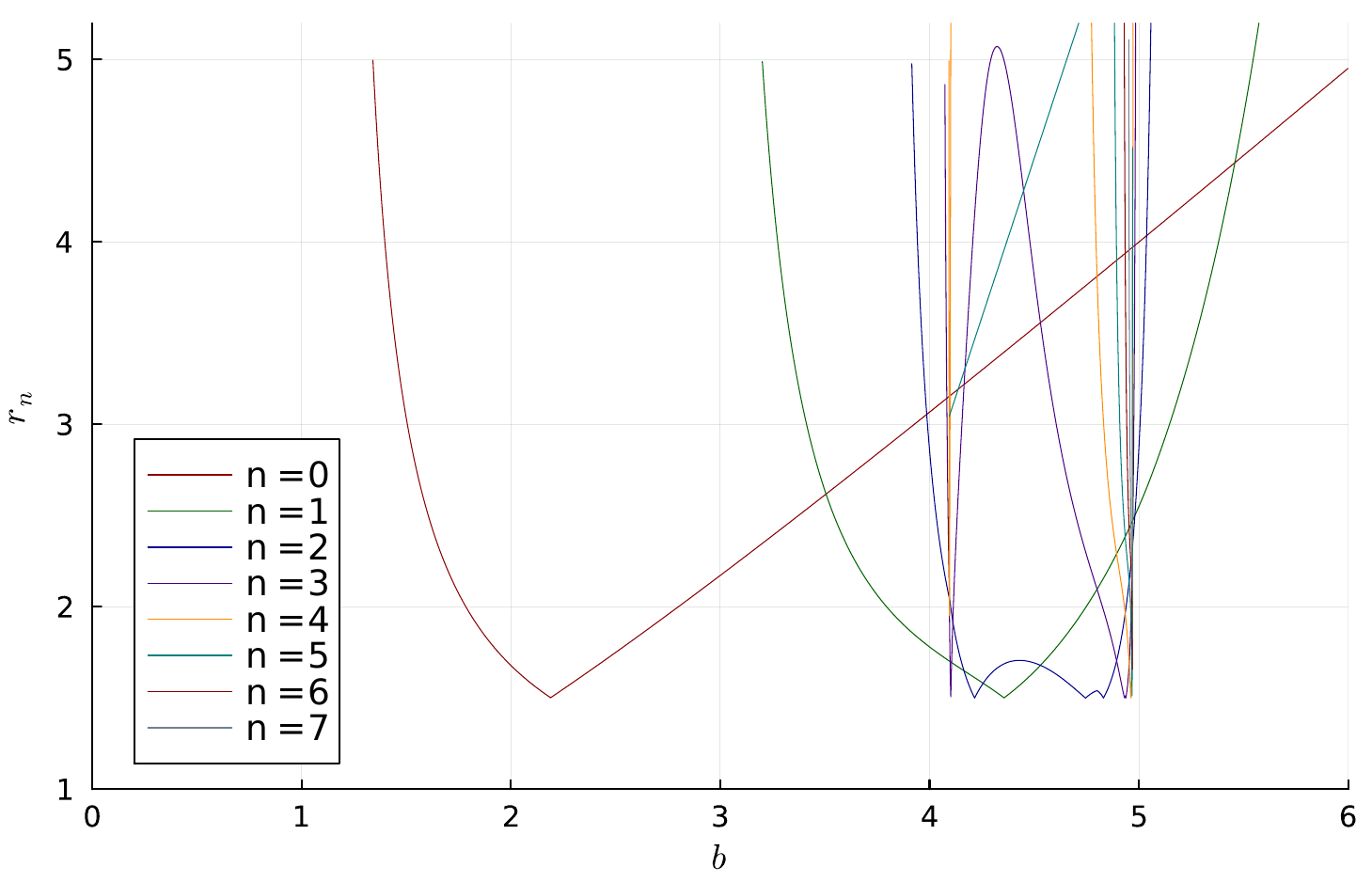} 
    \end{tabular}
    \caption{Transfer function $r=r_n(b)$ as a function of the impact parameter. The left image corresponds to a Reissner-Nordstr\"{o}m black hole, the middle image to an asymmetric wormhole with a single accretion disk, and the right image to the two-accretion disks case. In the two wormhole cases we see the presence of eight curves corresponding to the $n=0,1,2,3,4,5,6,7$ contributions to the image. }
    \label{fig: transfer}
\end{figure*}

The generation of images in our setting consists of writing down the coordinate $r$ in which the photon reaches the accretion disk (i.e. the values of $r$ fulfilling Eq.(\ref{eq:ndef})) and calculate the emission intensity of the disk on that point. The natural number $n$ represents the number of half-turns performed by the photon around the wormhole following Eq.(\ref{eq:ndef}) and where the factor $\pi/2$ accounts for the fact that a non-deflected photon from and to asymptotic infinity turns such an angle. As stated above, while for black holes it is sufficient to consider $n=0,1,2$, this no longer applies to the one-disk wormhole case, and for the numerical integration we shall consider $n=0,1,2,3,4,5,6,7$. The integration of Eq.\eqref{eq: geodesic} is done differently in the three sub-cases discussed in the ray-tracing, as described below.

\begin{itemize}

\item $b<b_c^-$. We consider the following initial conditions  $r_i = \sqrt{x_i^2+b^2}$, $\phi_0 = \arctan\frac{b}{x_i}$  and $r_f=r_0$, and we integrate the geodesic equation in $\mathcal{M}_+$ as an ingoing trajectory as
\begin{equation}
   \phi_+^I(r) = \int_{r_i}^{r} \frac{b}{r^2\sqrt{1-b^2\frac{A_+(r)}{r^2}}}dr + \phi_0,
\end{equation}
and the integration is carried from $r_i$ to $r_f$.  The photon then traverses to the $\mathcal{M}_-$ side  with an outgoing geodesic with the initial condition $\phi_1=\phi_+(r_f)$,
\begin{equation}
   \phi_-^O(r) = -\int_{r_f}^{r} \frac{b}{r^2\sqrt{1-b^2\frac{A_-(r)}{r^2}}}dr + \phi_1.
\end{equation}

\item $b_c^- \leq b\leq b_c^+$. We consider the following initial conditions  $r_i = \sqrt{x_i^2+b^2}$, $\phi_0 = \arctan\frac{b}{x_i}$  and $r_f=r_0$, and integrate the geodesic equation
\begin{equation}
   \phi_+(r) = \int_{r_i}^{r} \frac{b}{r^2\sqrt{1-b^2\frac{A_+(r)}{r^2}}}dr + \phi_0.
\end{equation}
The photon crosses the throat to $\mathcal{M}_-$ and keeps its path with the new initial condition $\phi_1 = \phi_+(r_f)$, $r^-_i =r_0$ and $r^-_f$ such that $V_-(r^-_f)-1/b=0$. The integral then becomes
\begin{equation}
   \phi_-(r) = -\int_{r^-_i}^{r} \frac{b}{r^2\sqrt{1-b^2\frac{A_-(r)}{r^2}}}dr + \phi_1.
\end{equation}
Next the photon turns back in $\mathcal{M}_-$ with the initial condition $\phi_2 = \phi_-(r_f)$, and the geodesic integral is now
\begin{equation}
   \phi_-(r) = \int_{r^-_f}^{r} \frac{b}{r^2\sqrt{1-b^2\frac{A_-(r)}{r^2}}}dr + \phi_2.
\end{equation}
Finally we have an outgoing geodesic in $\mathcal{M}_+$ with $\phi_3=\phi_-(r_i)$ and the integral
\begin{equation}
   \phi_+(r) = -\int_{r_f}^{r} \frac{b}{r^2\sqrt{1-b^2\frac{A_+(r)}{r^2}}}dr + \phi_3.
\end{equation}

\item  $b>b_c^+$. The photon stays always in $\mathcal{M}_+$. We consider the initial conditions  $r_i = \sqrt{x_i^2+b^2}$, $\phi_0 = \arctan\frac{b}{x_i}$  and $r_f$ such that $V_+(r_f)-1/b = 0$, and we integrate the geodesic equation  
\begin{equation}
   \phi_+(r) = \int_{r_i}^{r} \frac{b}{r^2\sqrt{1-b^2\frac{A_+(r)}{r^2}}}dr + \phi_0.
\end{equation}
The photon reaches $r_f$ and turns around, so we have an outgoing geodesic with a new initial condition $\phi_1 = \phi_+(r_f)$. The geodesic integral becomes
\begin{equation}
   \phi_+(r) =- \int_{r_f}^{r} \frac{b}{r^2\sqrt{1-b^2\frac{A_+(r)}{r^2}}}dr + \phi_1.
\end{equation}

\end{itemize}

\begin{figure*}[t!]
    \centering
    \begin{tabular}{ccc}
        \includegraphics[width=0.3\textwidth]{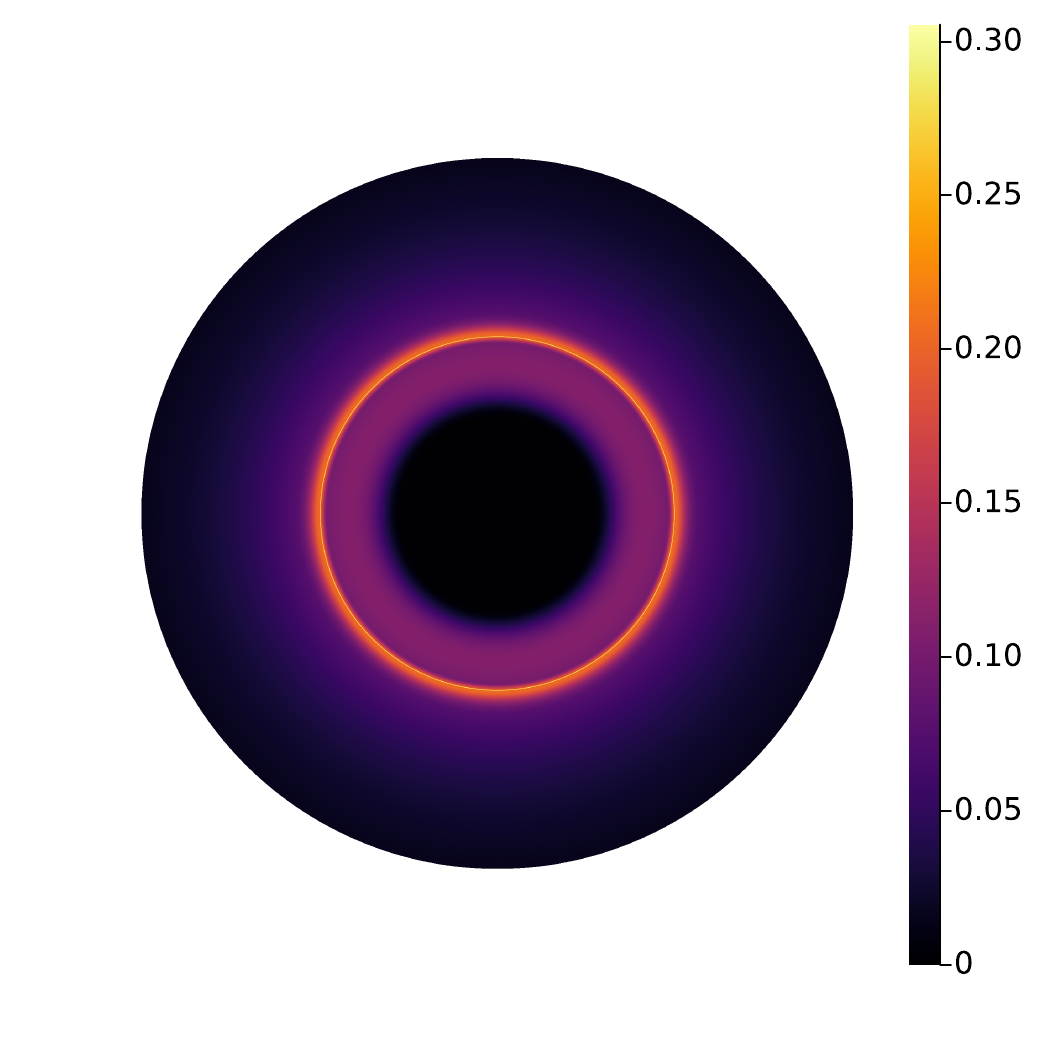} &
        \includegraphics[width=0.3\textwidth]{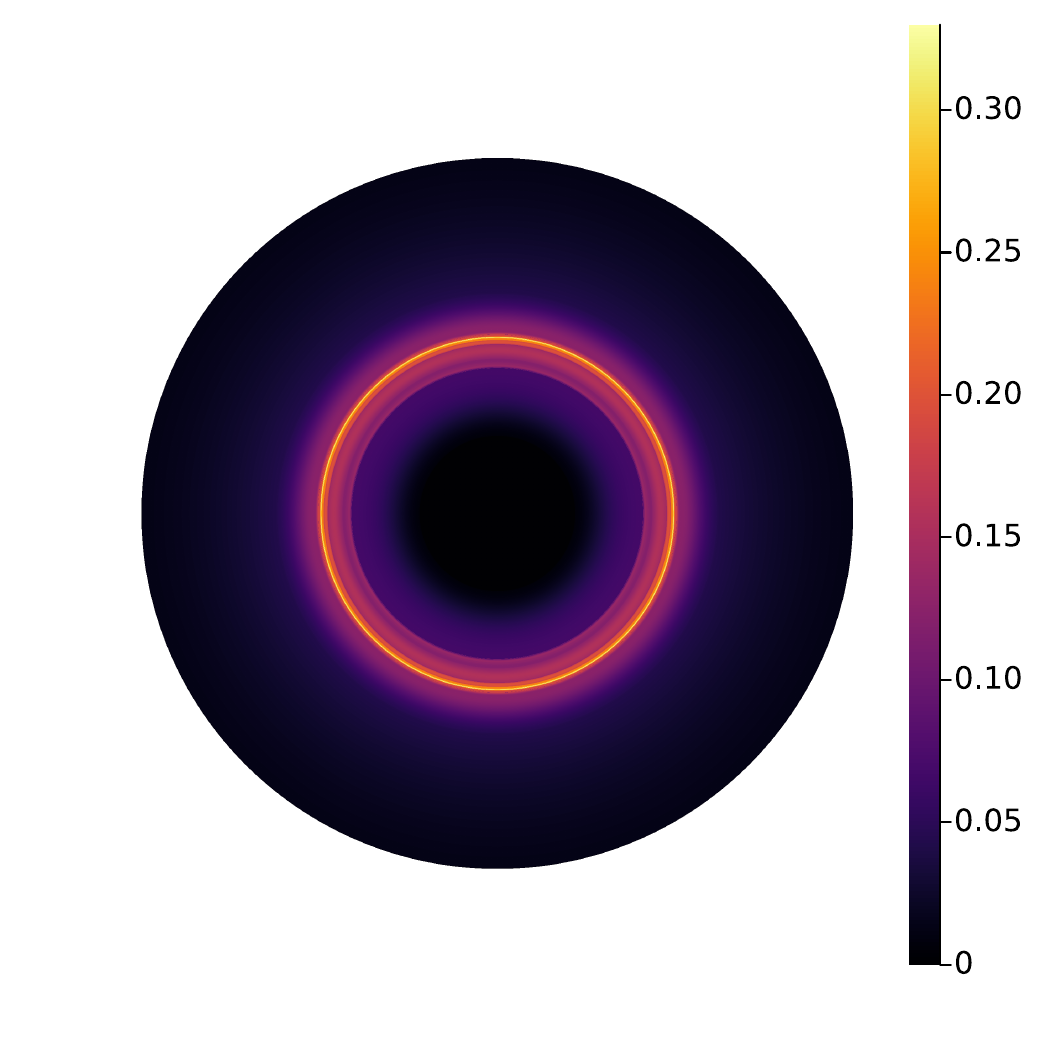} &
        \includegraphics[width=0.3\textwidth]{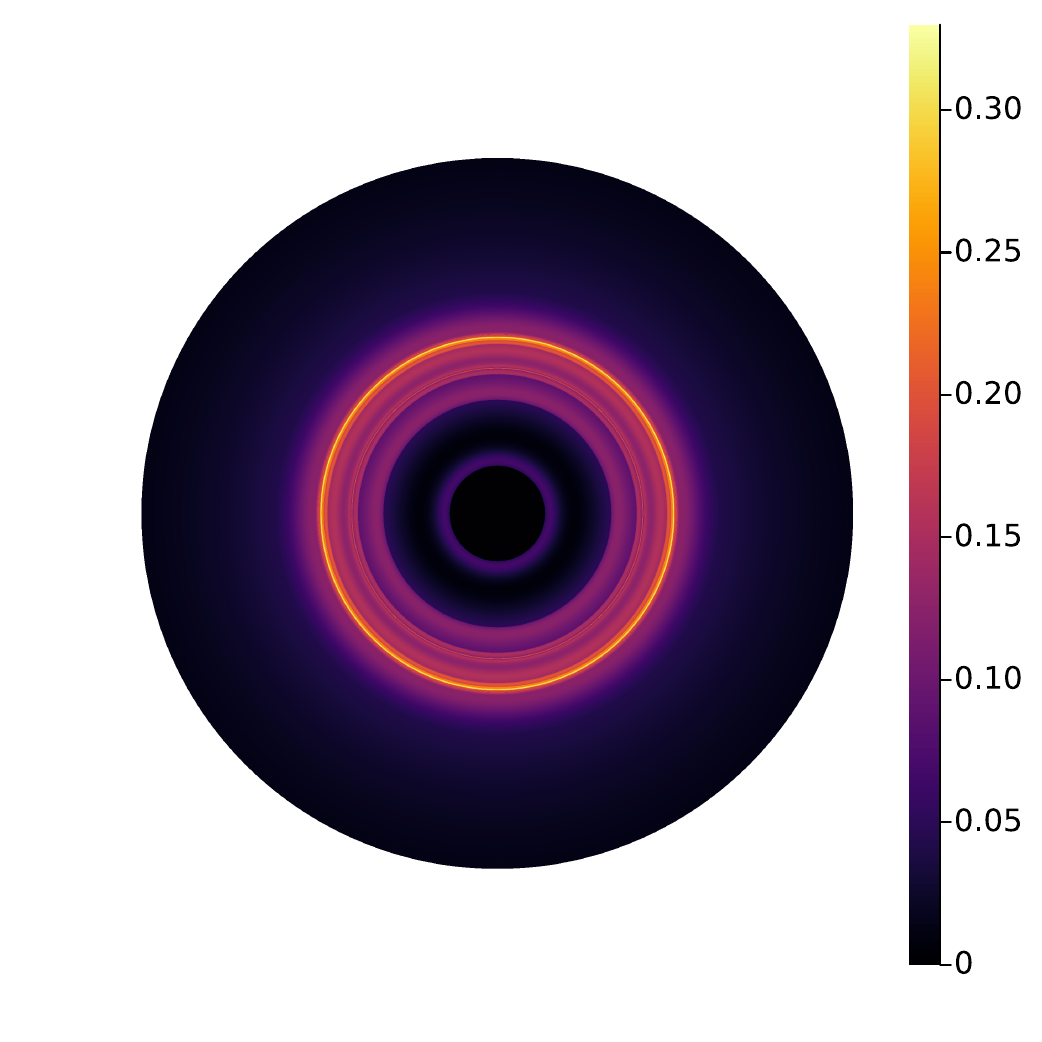} \\
        \includegraphics[width=0.3\textwidth]{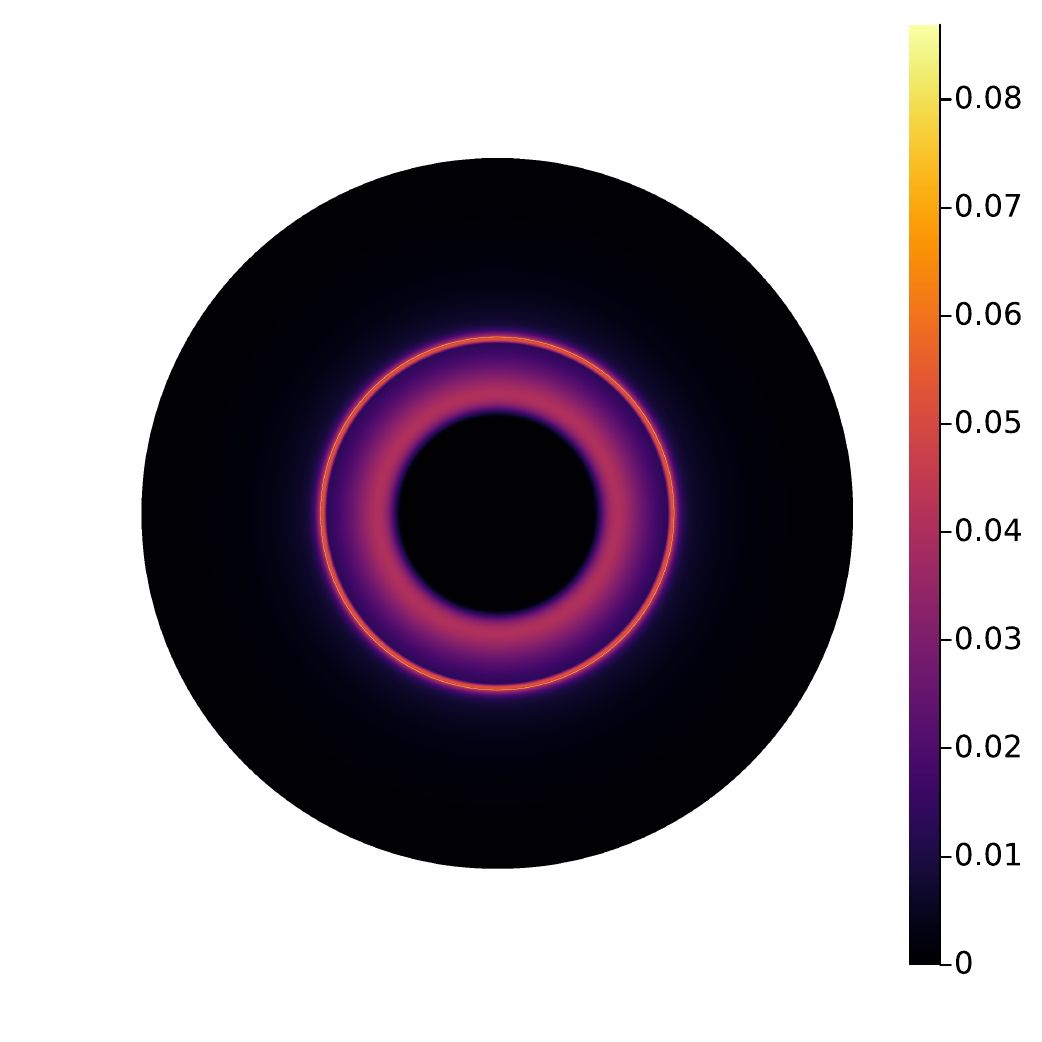} &
        \includegraphics[width=0.3\textwidth]{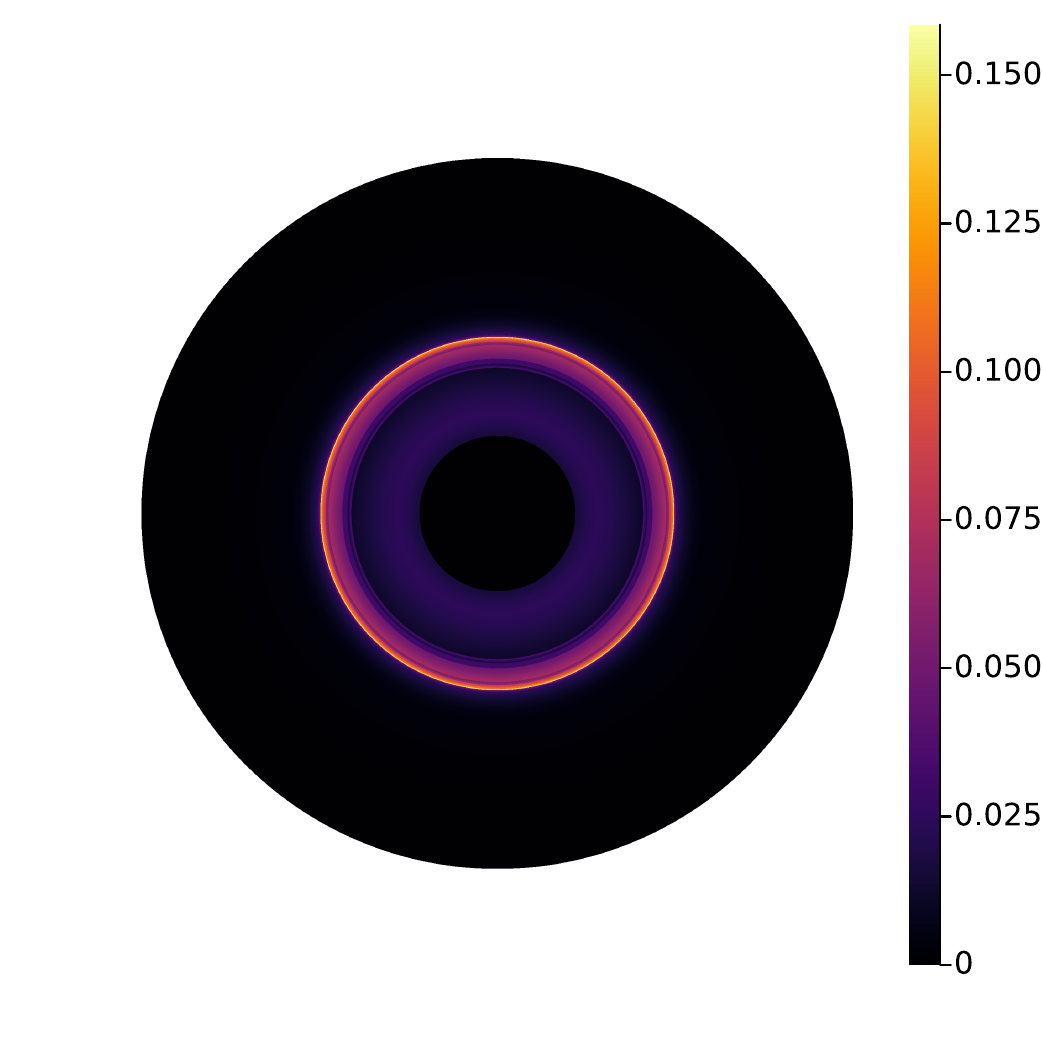} &
        \includegraphics[width=0.3\textwidth]{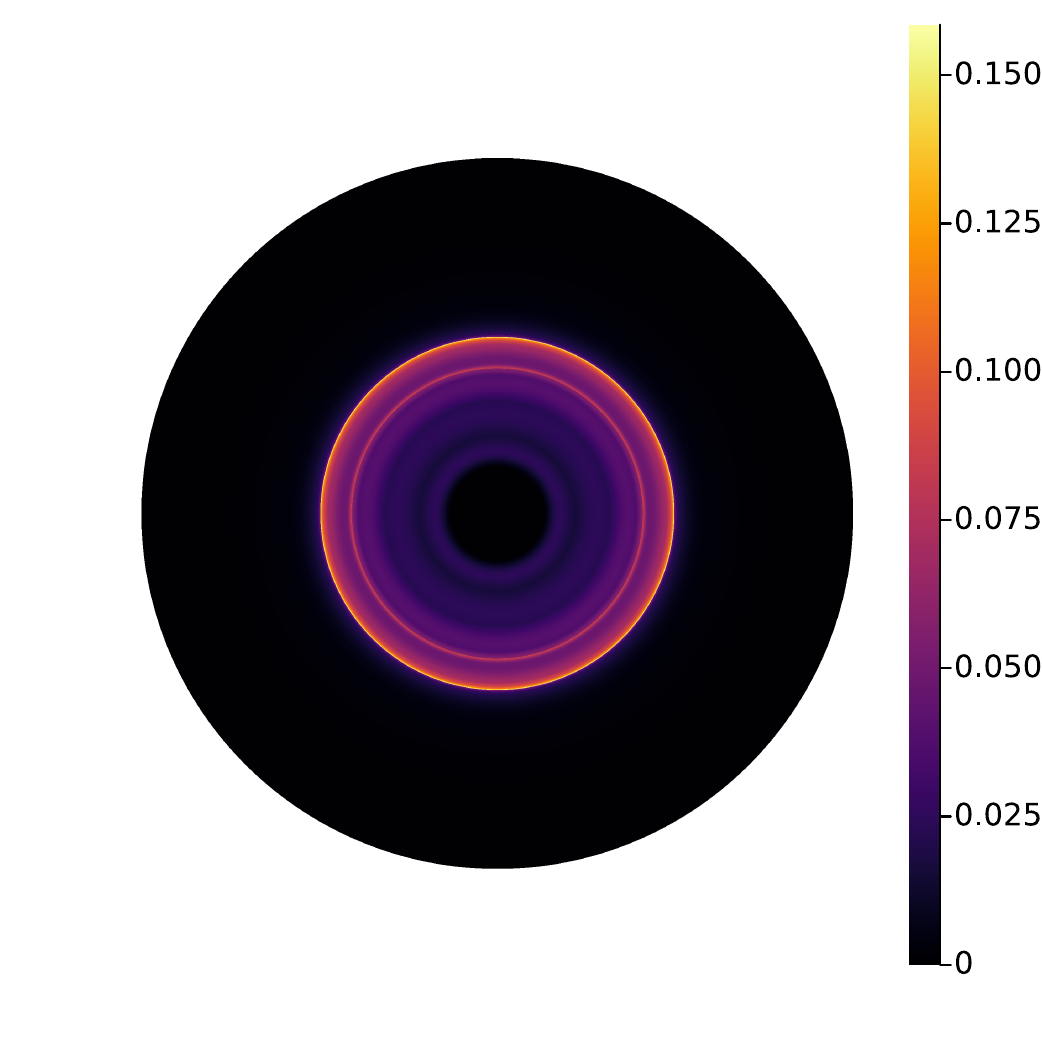} \\
        \includegraphics[width=0.3\textwidth]{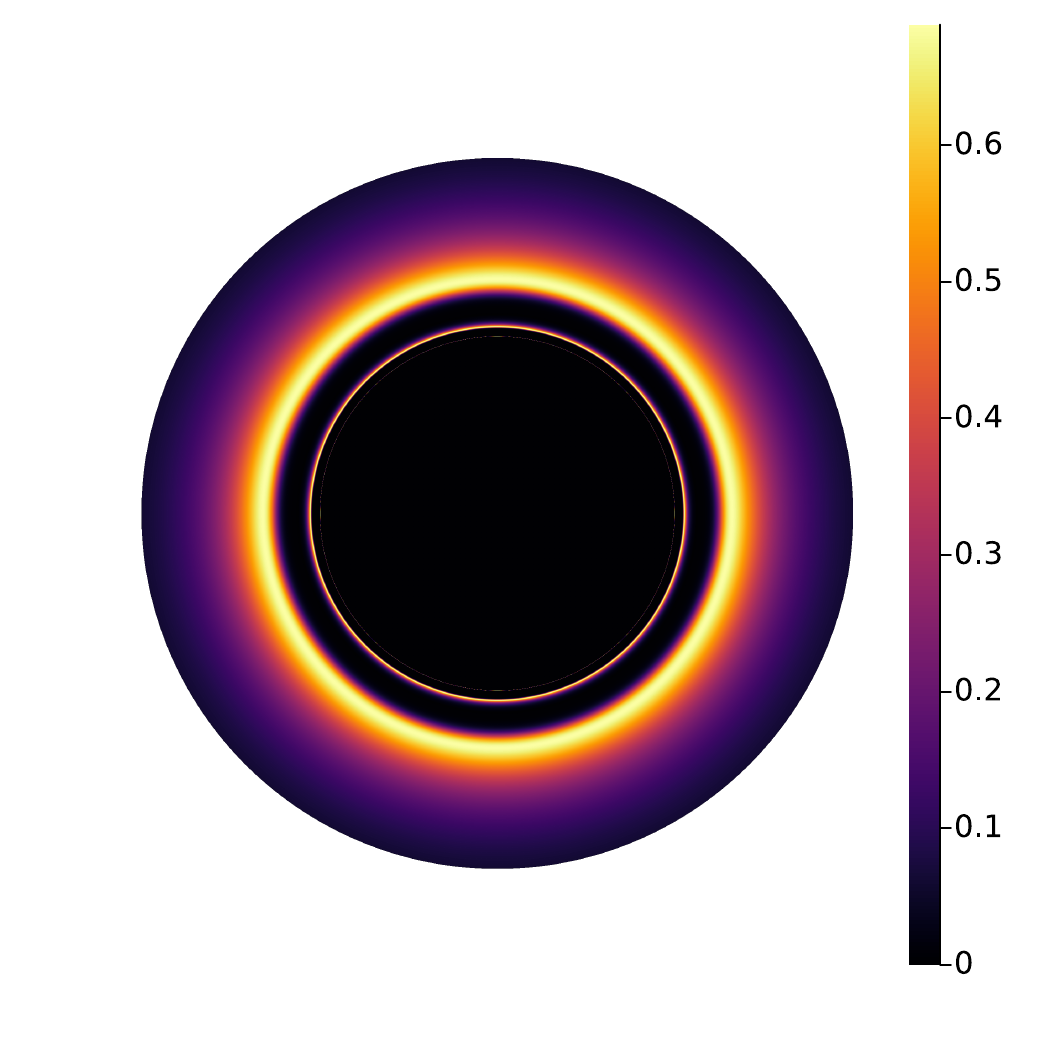} &
        \includegraphics[width=0.3\textwidth]{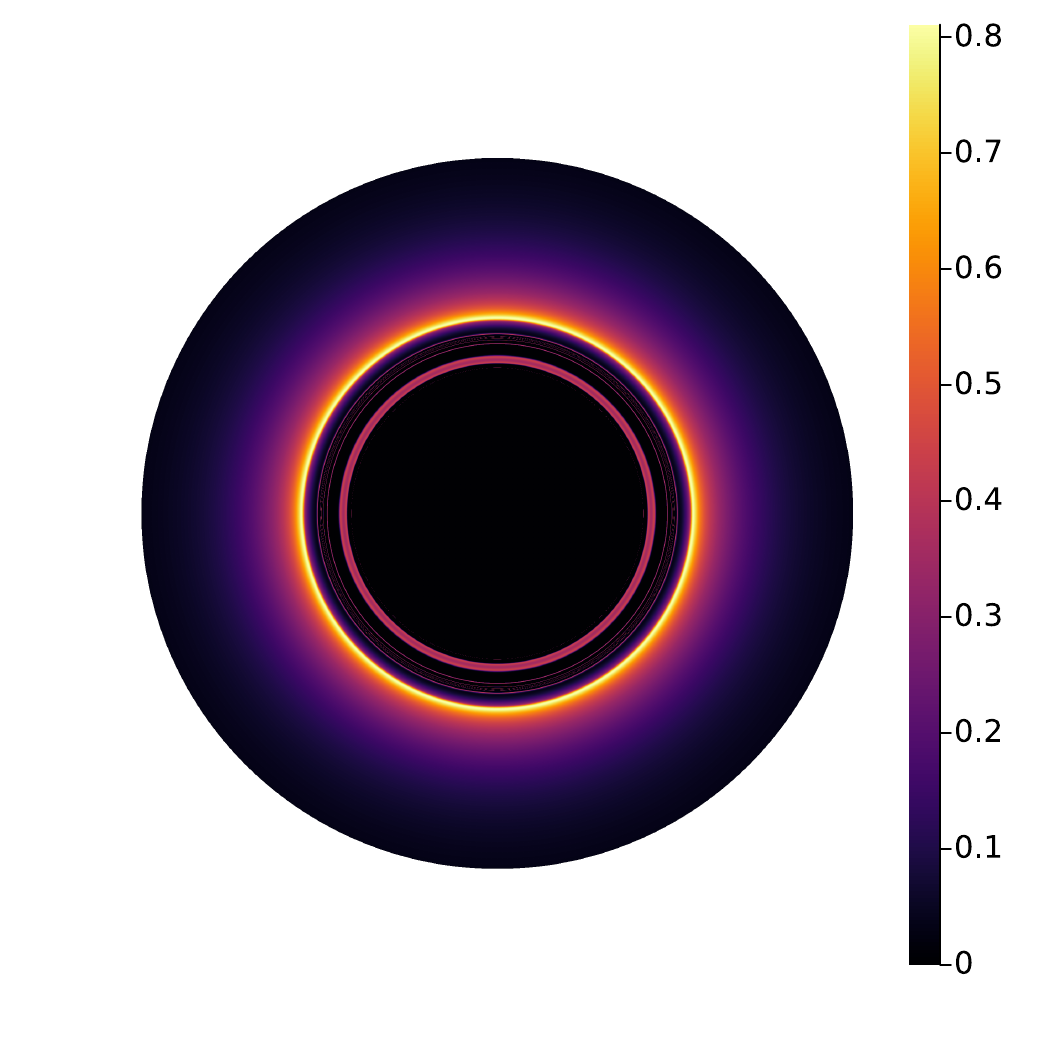} &
        \includegraphics[width=0.3\textwidth]{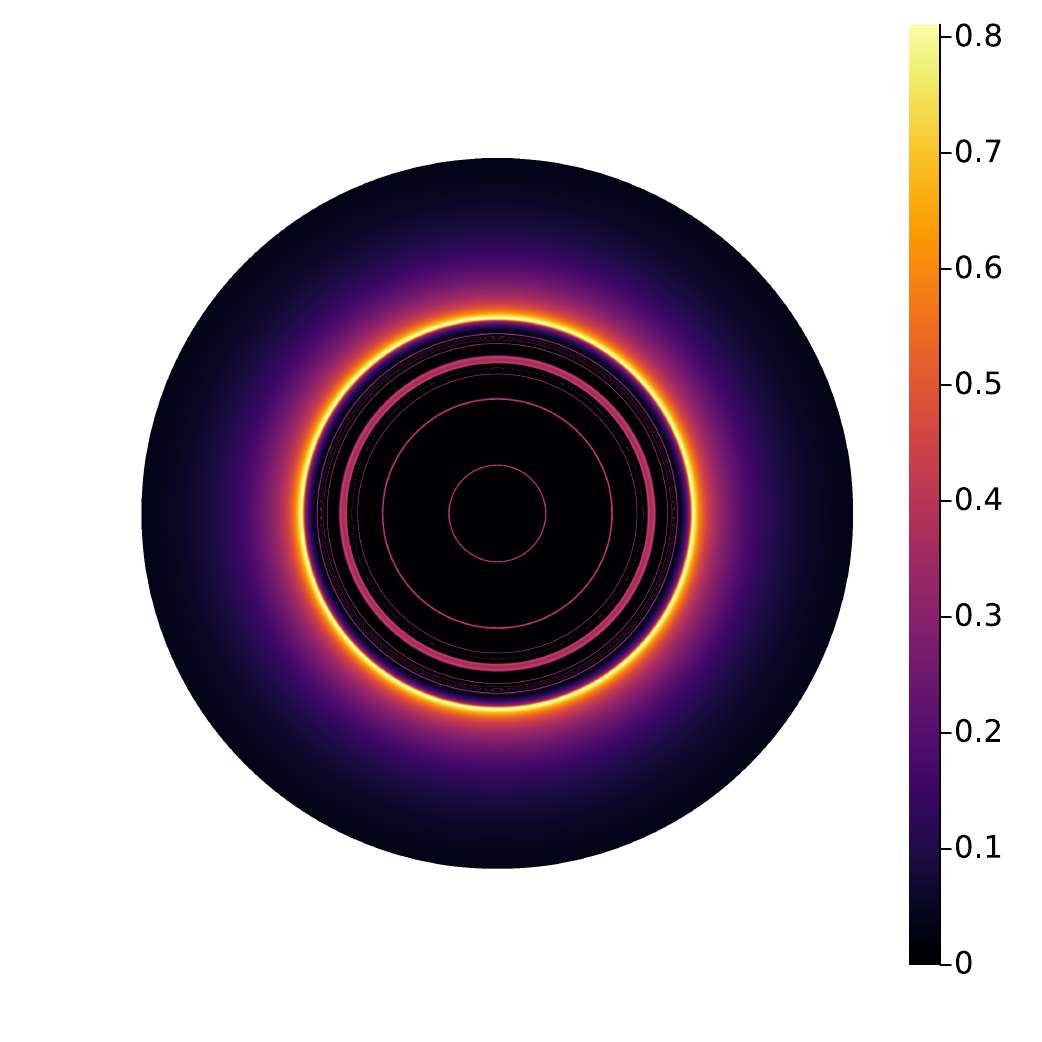} \\
    \end{tabular}
    \caption{Axial observational images. The left column corresponds to a Reissner-Nordstr\"om black hole, the middle column to an asymmetric wormhole with a single accretion disk, and the right column to two-accretion disks case. The top row corresponds to the intensity distribution GML1, the middle row to GML2, and the bottom row to GML3. Model parameters and units are those discussed in  Sec. \ref{sec:asym}.}
    \label{fig:map}
\end{figure*}

Following this procedure for the different scenarios we display in Fig. \ref{fig: transfer}  the transfer function for the usual Reissner-Nordstr\"om black hole (left), an asymmetric wormhole with a disk on one side (middle), and with disks on both sides (right). In the black hole case we find the typical curves associated to the direct emission ($n=0$) and to the first and second photon rings, with their slopes telling us about the relative degree of demagnification in the corresponding images \cite{Perlick:2021aok}. For the wormhole case, however, there are far more contributions due to the presence of the additional photon rings and, furthermore, we see that several of them have their ends taking a vertical slope. This is associated with those light trajectories that circulate near the two maxima of the effective potential. One should point out that additional, higher-order photon rings are difficult to track given the large sensibility of light trajectories nearing the corresponding critical impact parameters, which poses a challenge to numerical accuracy and efficiency.

\subsection{Analysis of the images: single thin disk on the observer's side}

\begin{figure*}[t!]
    \centering
    \begin{tabular}{ccc}
        \includegraphics[width=0.3\textwidth]{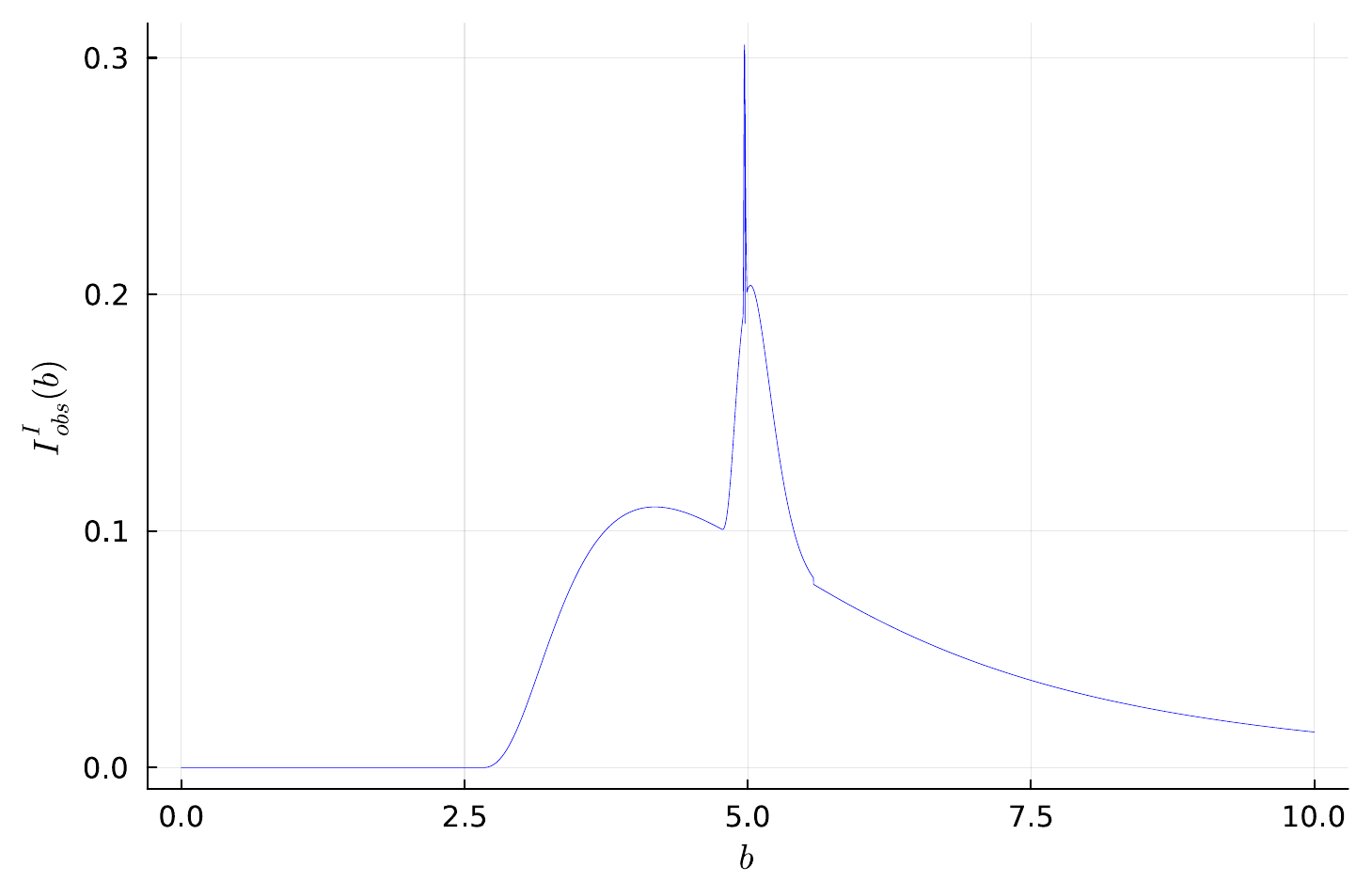} &
        \includegraphics[width=0.3\textwidth]{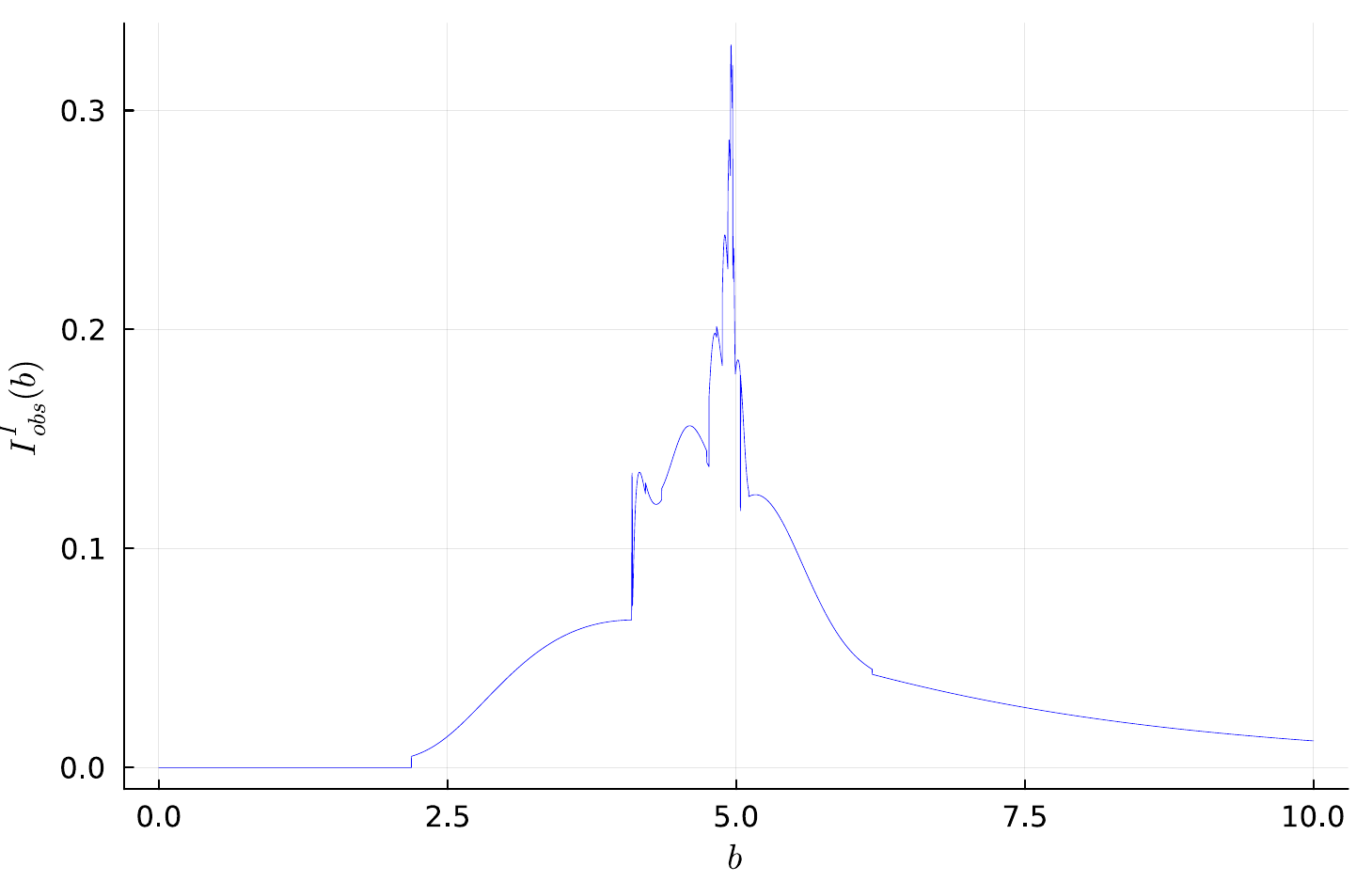} &
        \includegraphics[width=0.3\textwidth]{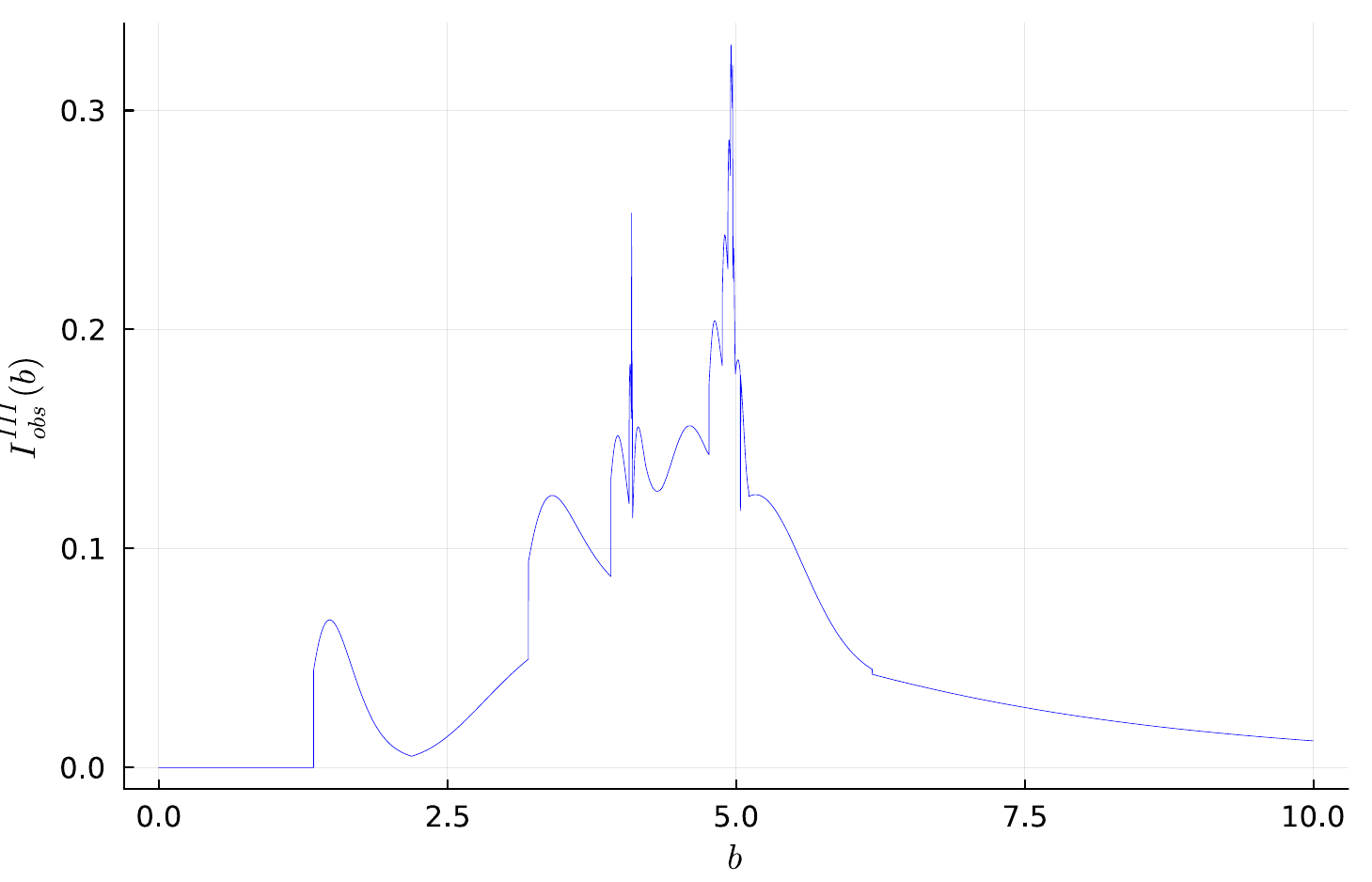} \\
        \includegraphics[width=0.3\textwidth]{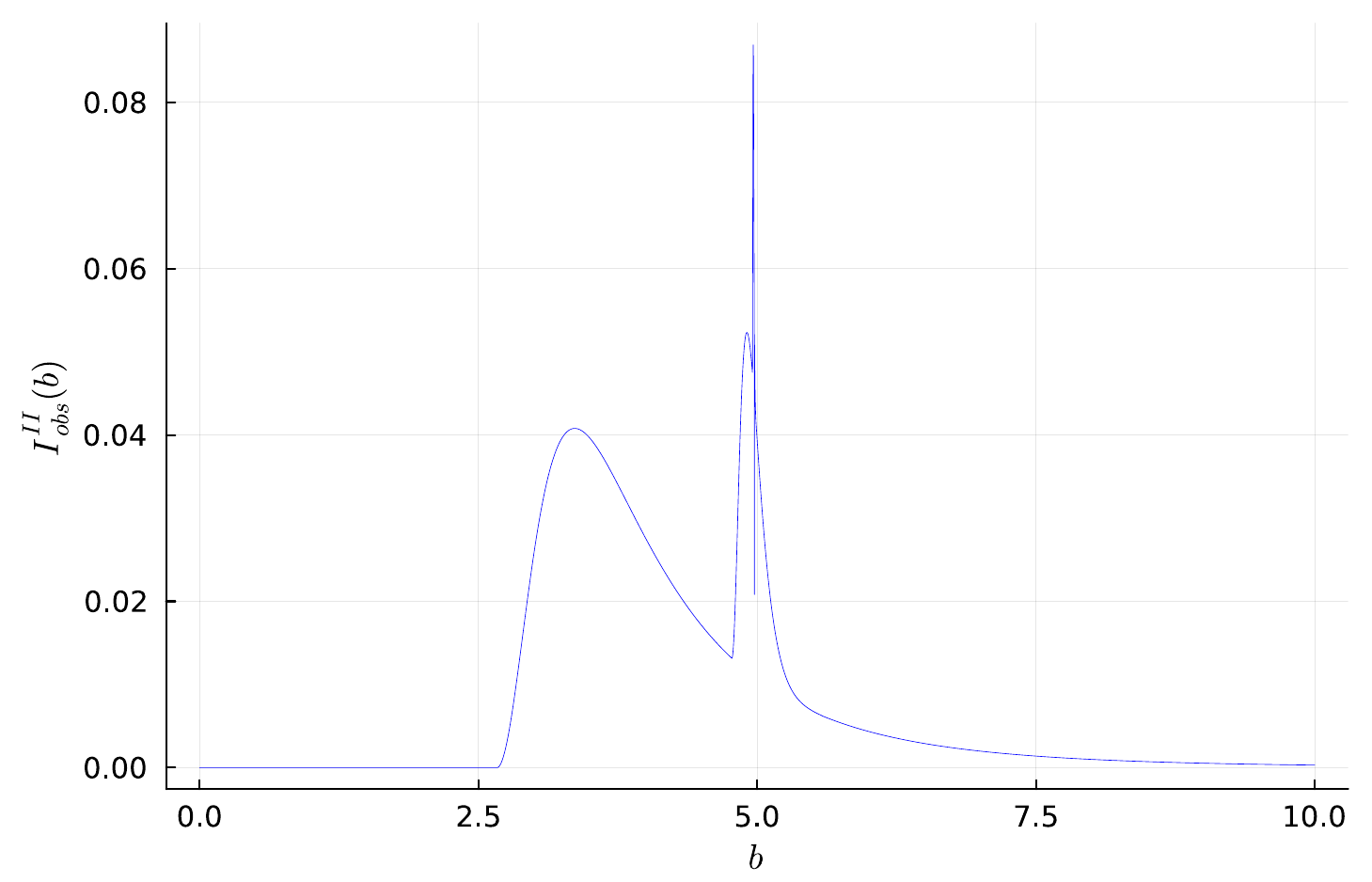} &
        \includegraphics[width=0.3\textwidth]{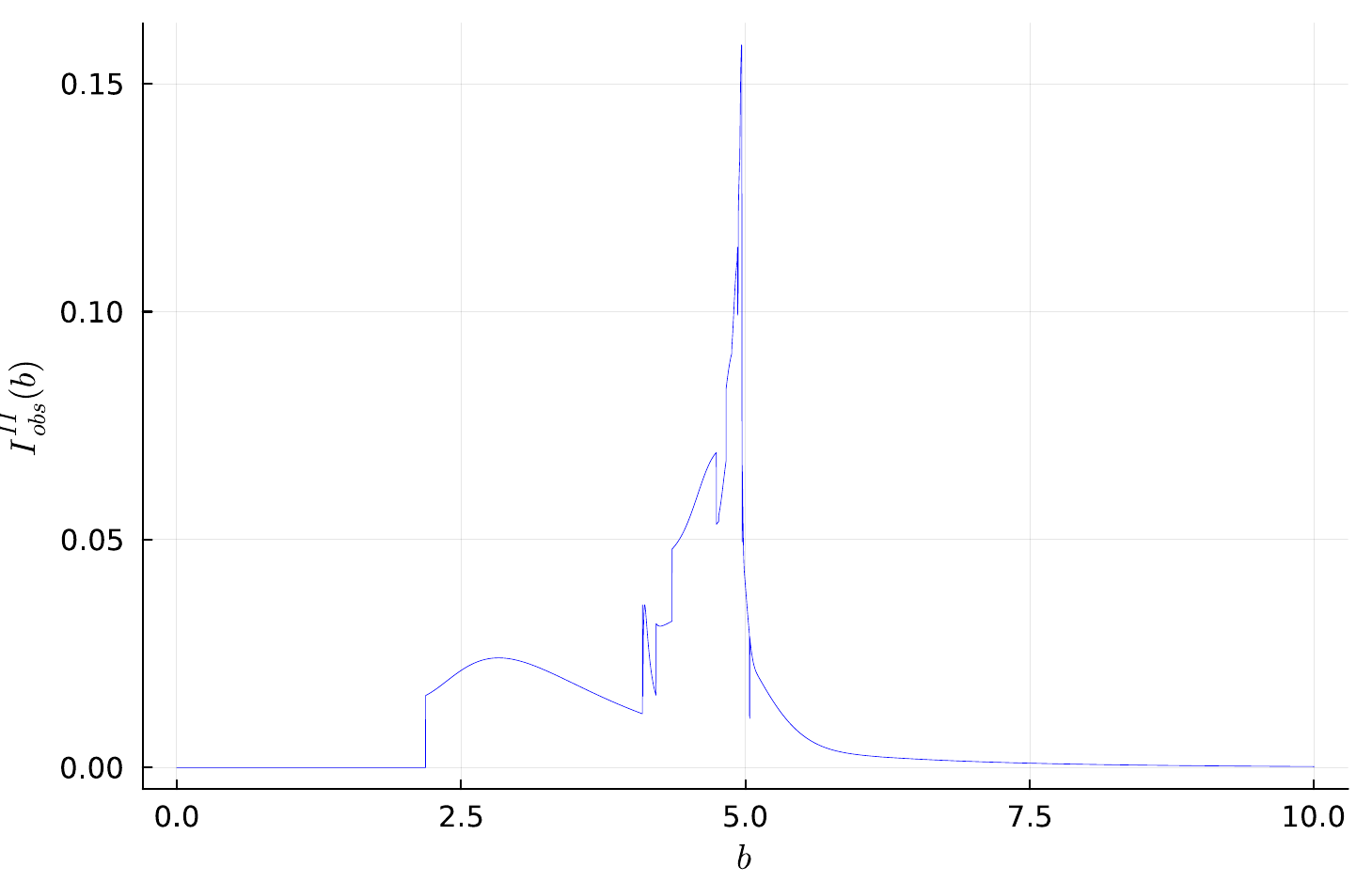} &
        \includegraphics[width=0.3\textwidth]{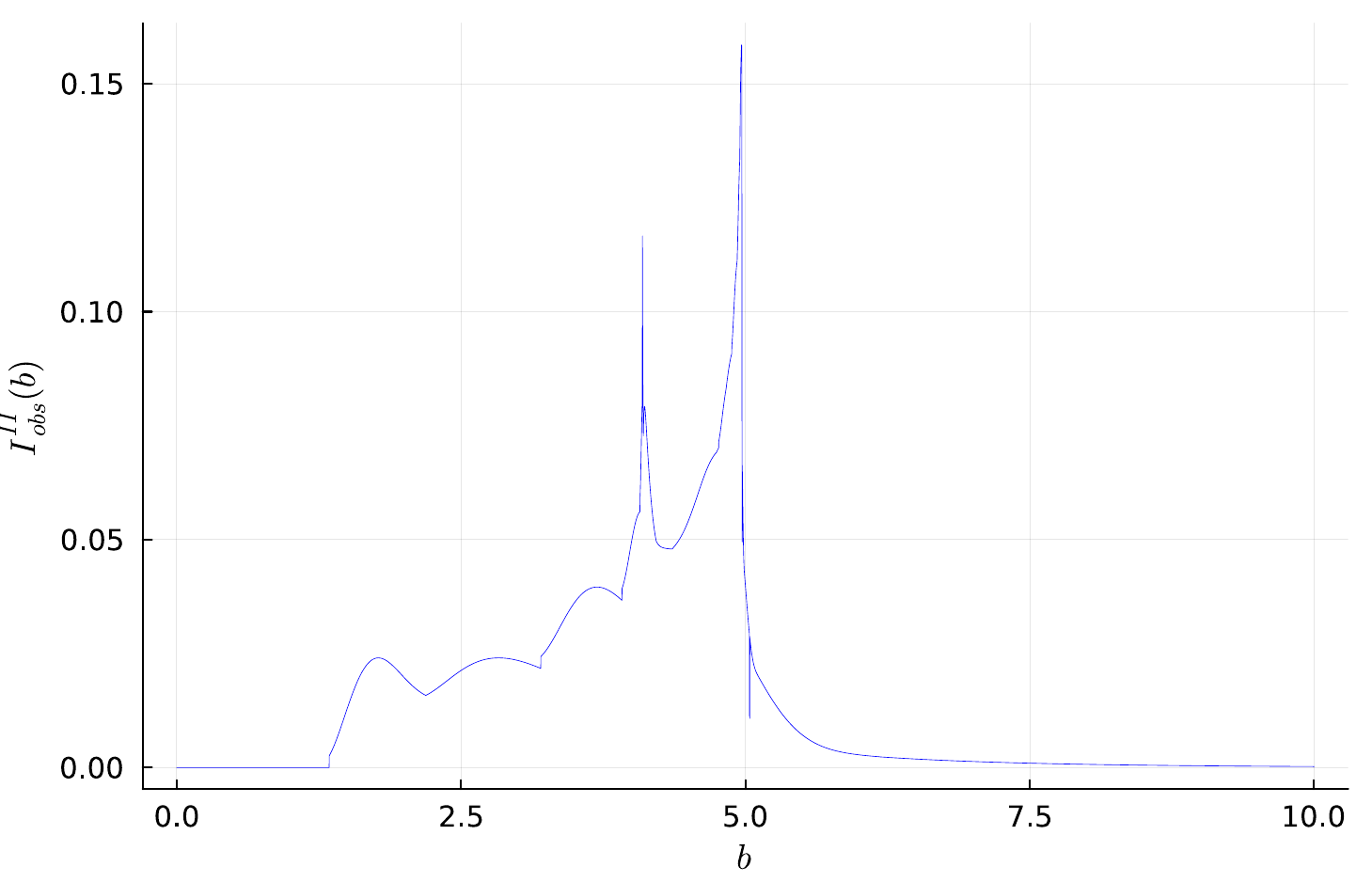} \\
        \includegraphics[width=0.3\textwidth]{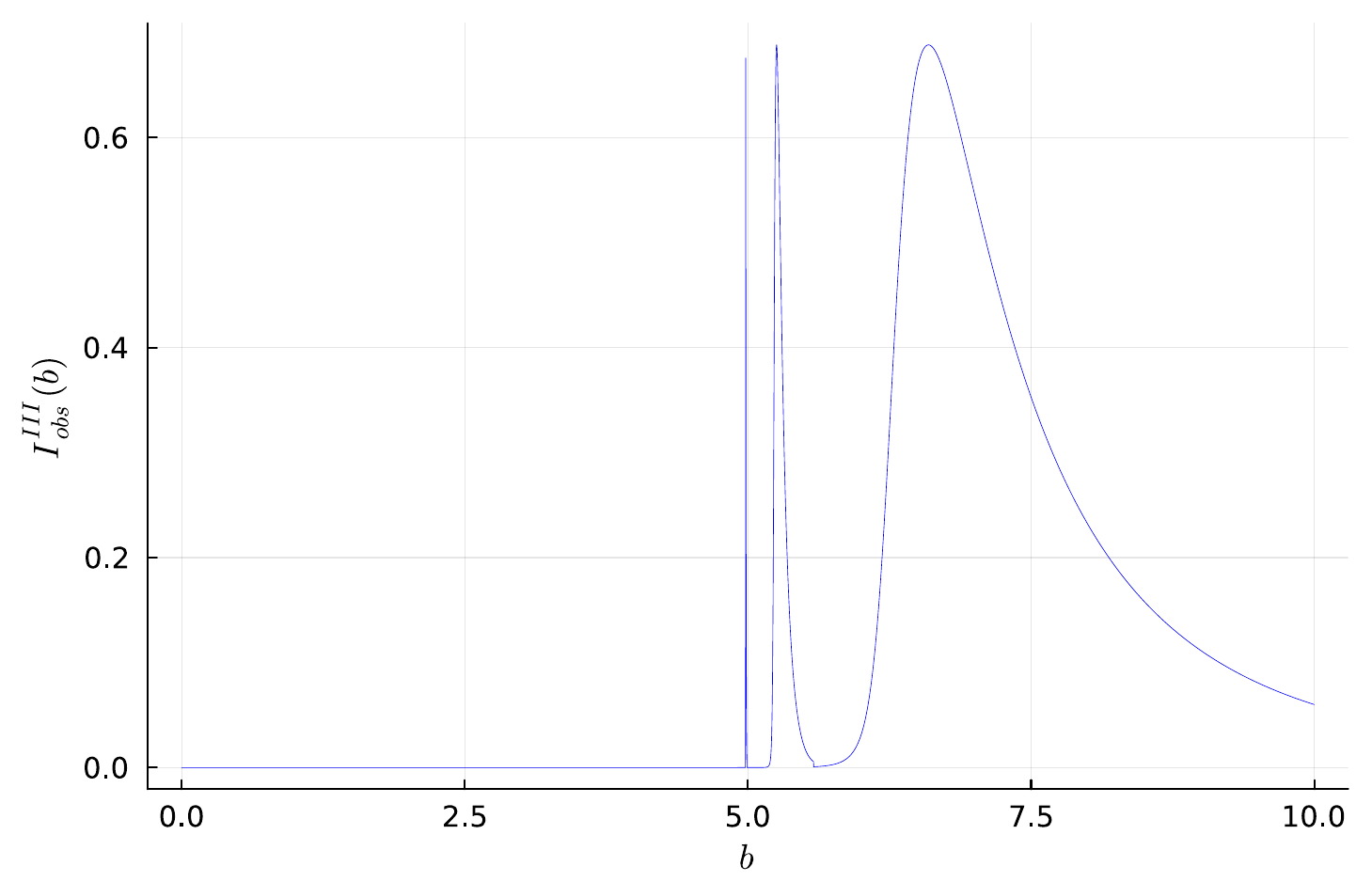} &
        \includegraphics[width=0.3\textwidth]{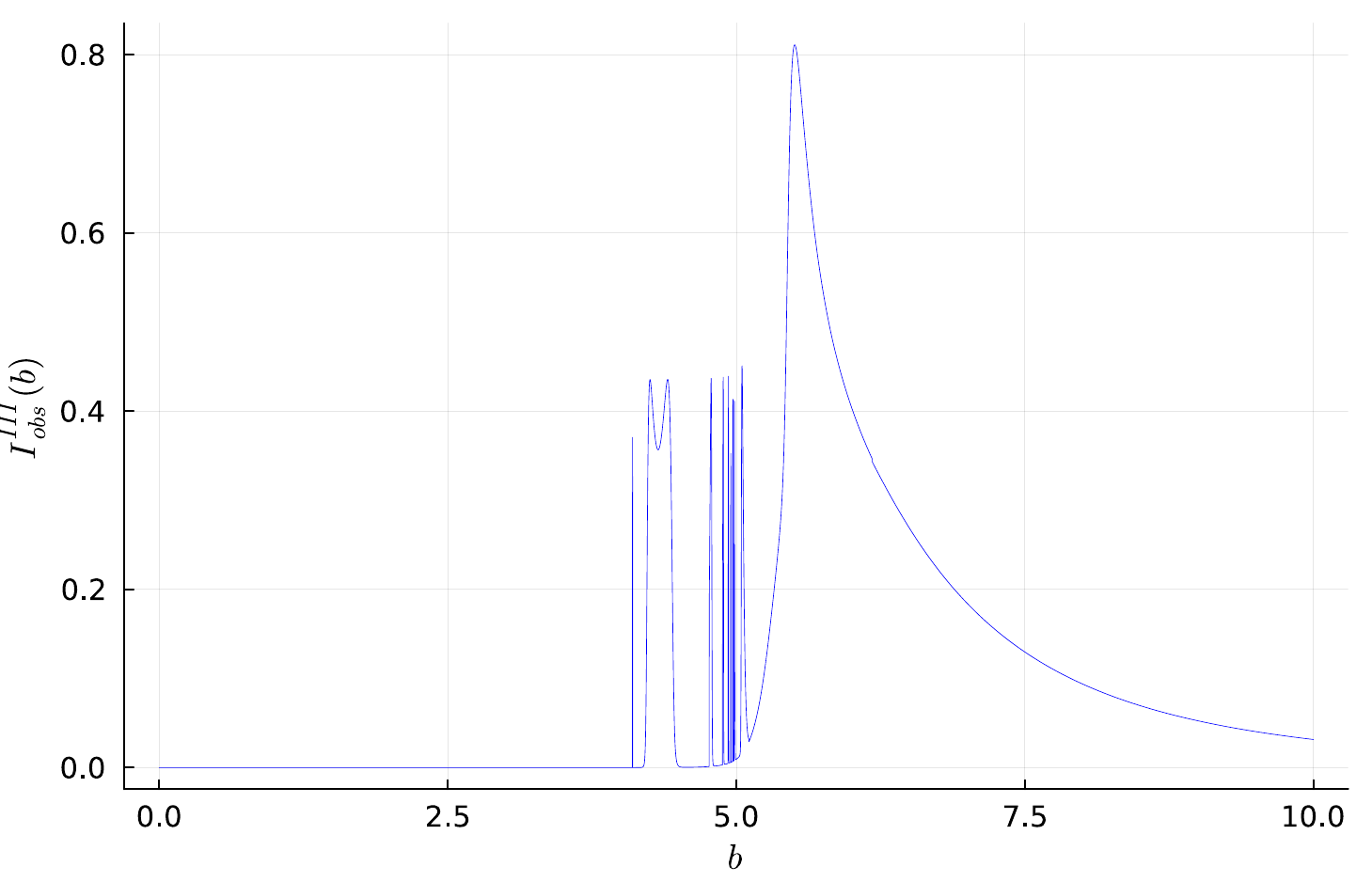} &
        \includegraphics[width=0.3\textwidth]{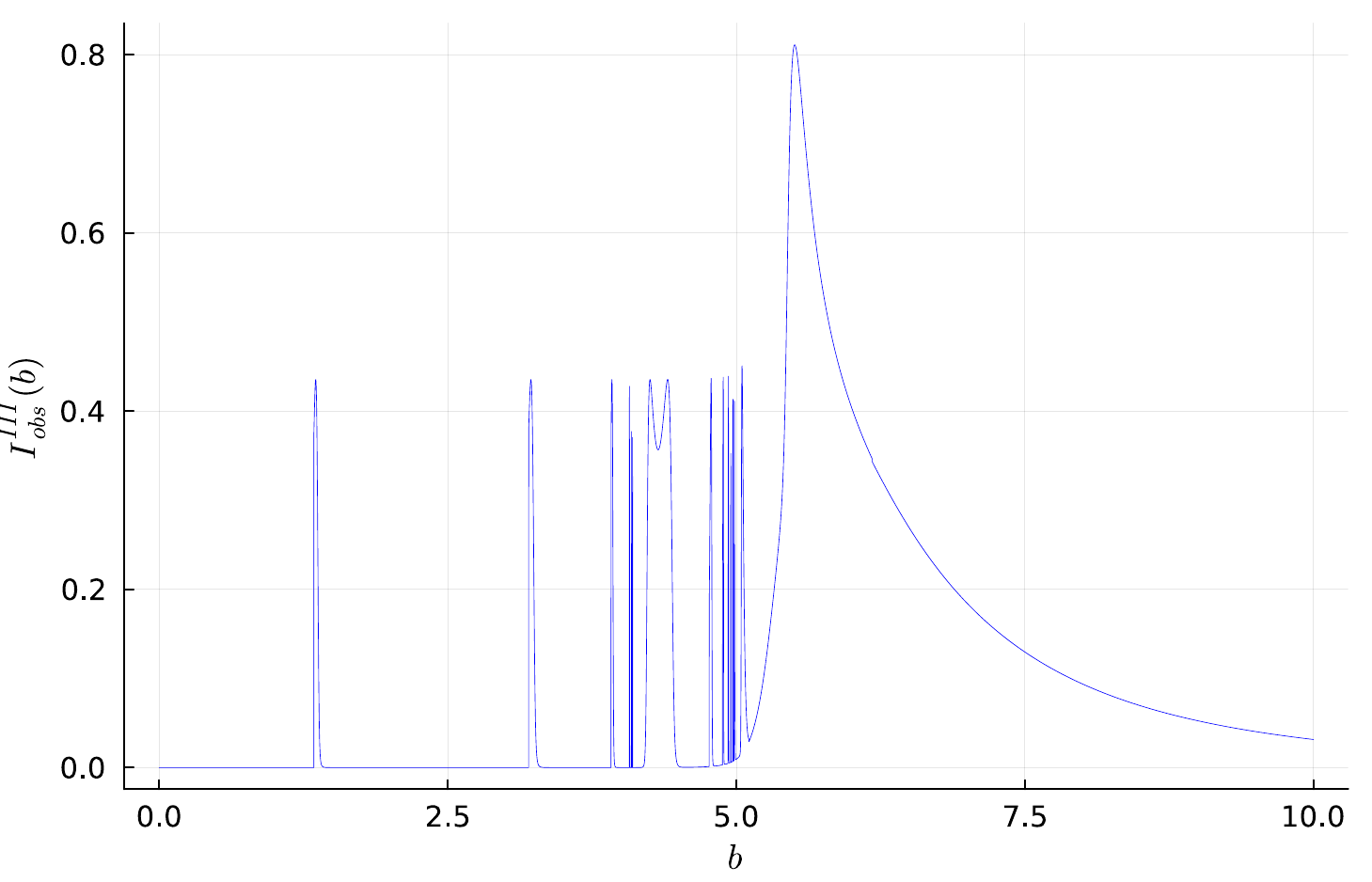} \\
    \end{tabular}
    \caption{Axial observed intensity as a function of the impact parameter. The left column corresponds to a Reissner-Nordstr\"om black hole, the middle column to an asymmetric wormhole with a single accretion disk, and the right column to a double accretion disk. The top row corresponds to the intensity distribution GML1, the middle row to GML2, and the bottom row to GML3.}
    \label{fig:obint}
\end{figure*}

Fig. \ref{fig:map} shows the optical appearance of axial images, namely, zero inclination between disk and observer, for the Reissner-Nordstr\"om black hole (left figures) the asymmetric wormhole with one accretion disk (middle figures), and the asymmetric wormhole with two disks (right figures), for the GLM1 (top figures), GLM2 (middle figures) and GLM3 (bottom figures) emission models.

Let us discuss the single-disk case first.

We see that there are significant differences between the optical appearances for all the GLM emission models. In every model, for the Reissner-Nordstr\"om black holes we find two sharp peaks of intensity alongside a far more spread intensity region. These peaks are associated with the  $n=1$ and $n=2$ photon rings, while the wide region is the contribution from direct emission, $n=0$. The latter contribution produces the typical wide ring of radiation enclosing a central brightness depression, whereas the photon rings appear as local insertions of luminosity inside the direct image (GLM1 and GLM2 models), or appear separately in the inner region to it (GLM3 model). In other words, while in the GLM1 and GLM2 models the effective shadow is determined by the direct emission, in the GLM3 model it is set instead by the innermost of the photon rings. This is a recurrent feature of black hole images regarding the correlation between the shadow's size and the effective peak of emission of its disk, up to a minimum theoretical size - the inner shadow \cite{Chael:2021rjo}.

As opposed to these black hole images, in the asymmetric traversable thin-shell wormhole, single-disk model we still have the direct emission dominating the optical appearance, but there are six additional photon rings, though barely visible in the image, which appear as further insertions of luminosity within the direct emission in the GLM1 and GLM2 models, and inner to it in the GLM3 model. These photon rings can be better visualized by appealing to the observed intensity profiles, which we depict in Fig. \ref{fig:obint} (corresponding again, from left to right, to the Reissner-Nordstr\"om black hole, the asymmetric wormhole with a single disk, and the case with two disks). In this figure we can neatly see the novelties the asymmetric wormhole scenario brings in: the observed intensity has a much more intricate structure, being decomposed into many thin but short peaks, which become the source of the additional photon rings in the full image. As discussed above, these additional photon rings arise due to the light rays that are able to traverse the throat, turn around near any of the photon spheres several times, and return to the observer's screen on the $\mathcal{M}_+$ side. This way, the structure of new higher-order rings becomes an observational discriminator of asymmetric wormholes as compared to canonical black hole images, particularly on the grounds that they do not follow the universal scaling laws of canonical photon rings. The new rings, however, are quite faint in comparison with the usual $n=1$ black hole ring. Given the limited resolution of current interferometric (EHT) devices, and even with the planned experiments, such as the ngEHT \cite{Johnson:2023ynn} and the Black Hole Explorer \cite{Johnson:2024ttr}, which aim to measure the $n=1$ ring (and possibly the $n=2$ too) the chances of observation of these rings are difficult to foresee.

\subsection{Two thin-disks scenario}

Let us now consider the optical appearance of the asymmetric wormhole with two identical thin accretion disks, located on each side of the throat $x_0$ (i.e. on $\mathcal{M}_\pm$) at the same distance from it. We follow the same procedure as in the previous single thin-disk scenario for the integration of outward/inward geodesics from the disk on the $\mathcal{M}_+$ side, to which we add those emitted inward from the disk on the $\mathcal{M}_-$ side. This way, we have to consider the intersection of the photons with both accretion disks whenever this is the case (i.e. in the impact parameter window in which this may happen). 

\begin{figure*}[t!]
    \centering
    \begin{tabular}{ccc}
        \includegraphics[width=0.3\textwidth]{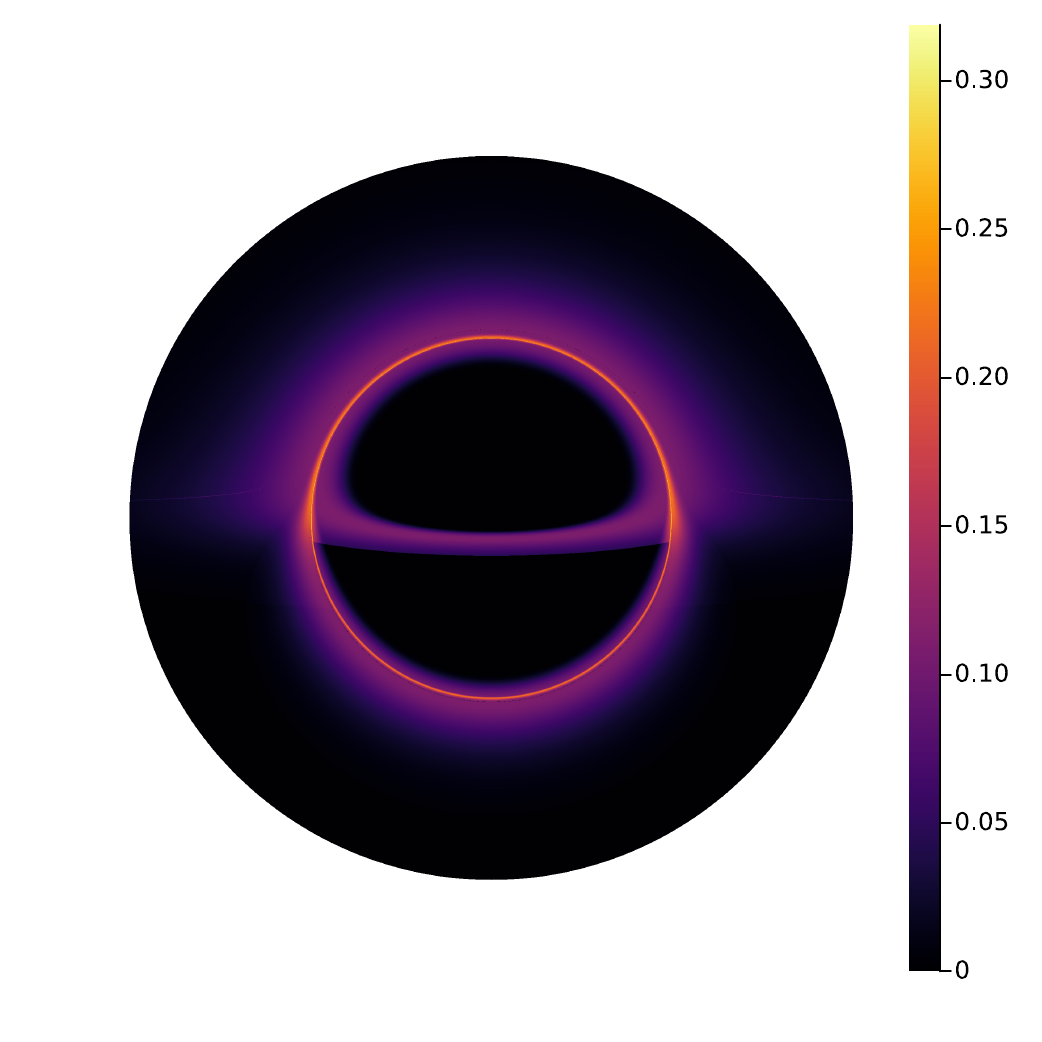} &
        \includegraphics[width=0.3\textwidth]{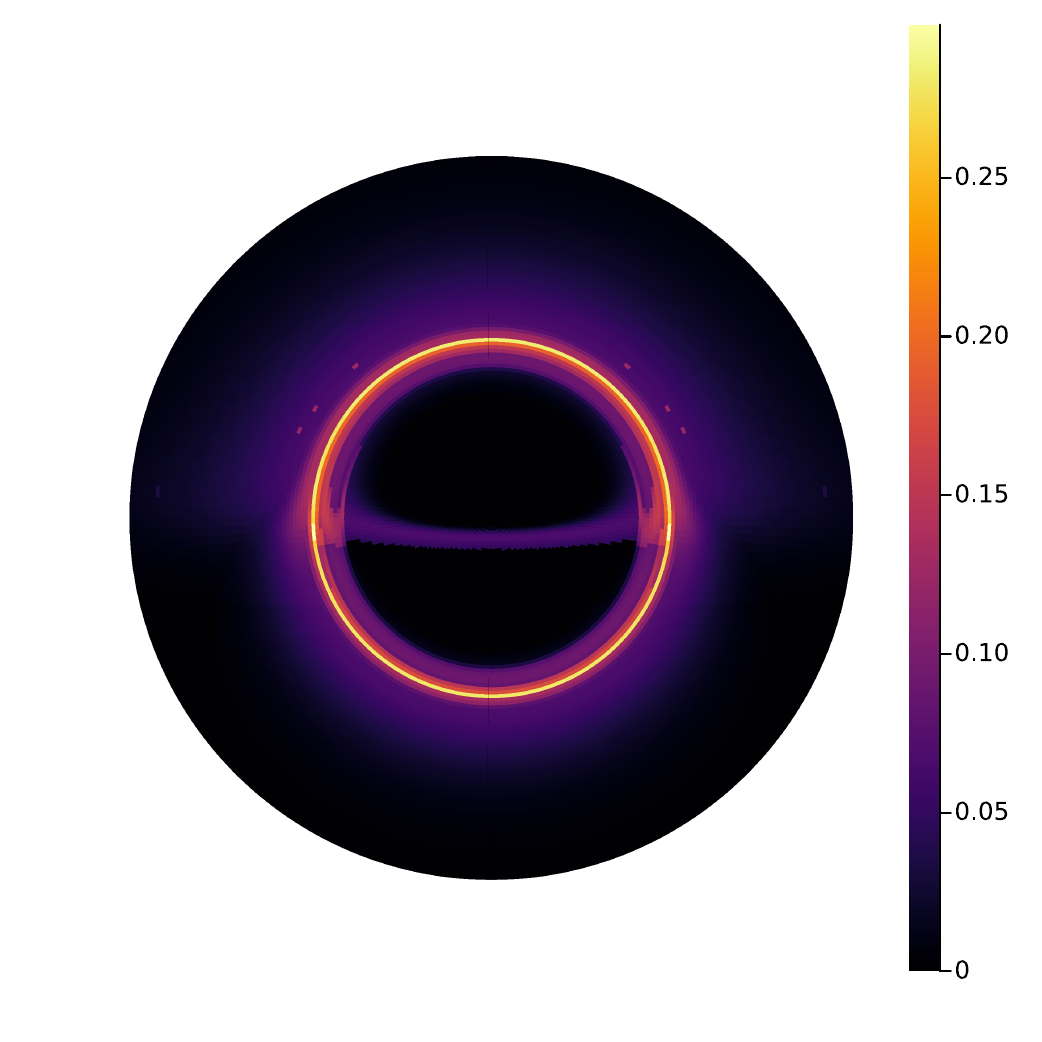} &
        \includegraphics[width=0.3\textwidth]{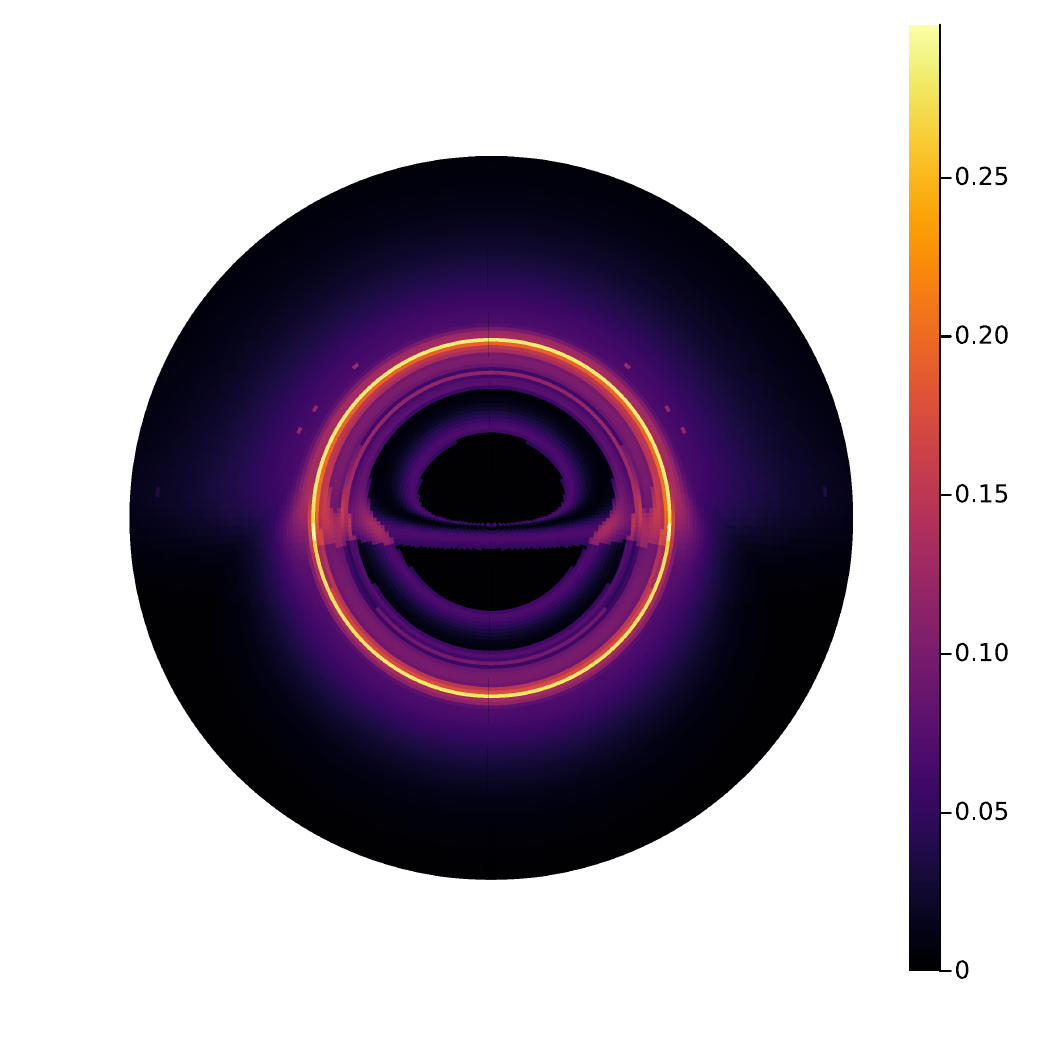} \\
        \includegraphics[width=0.3\textwidth]{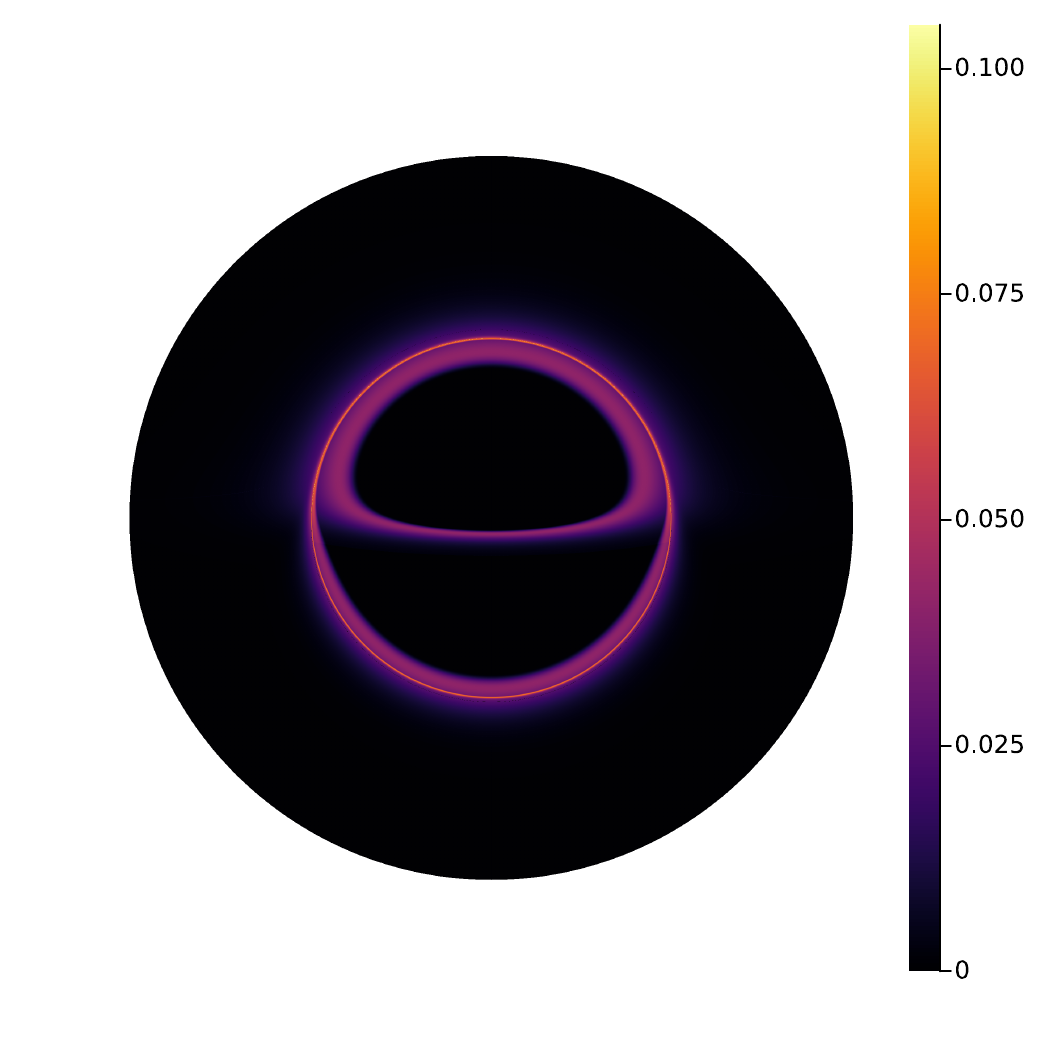} &
        \includegraphics[width=0.3\textwidth]{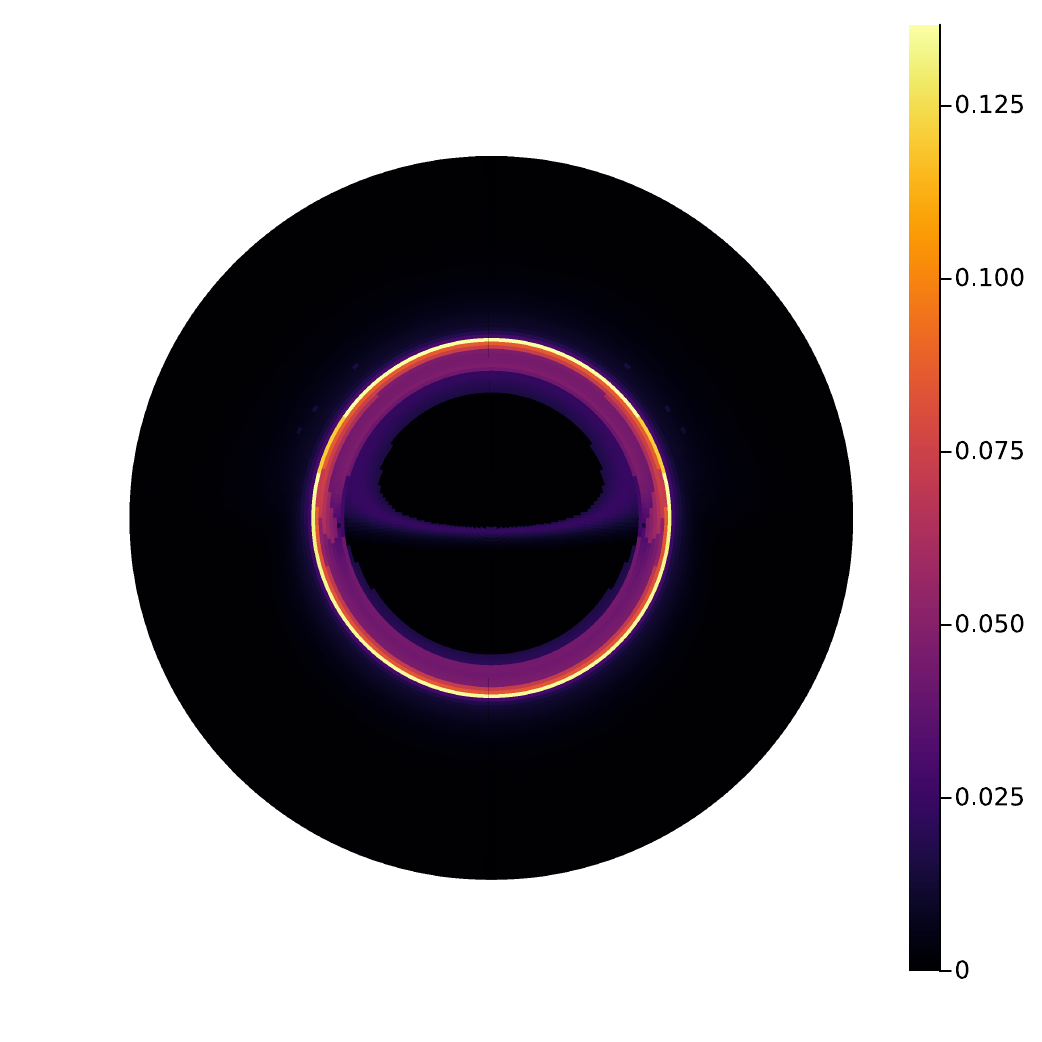} &
        \includegraphics[width=0.3\textwidth]{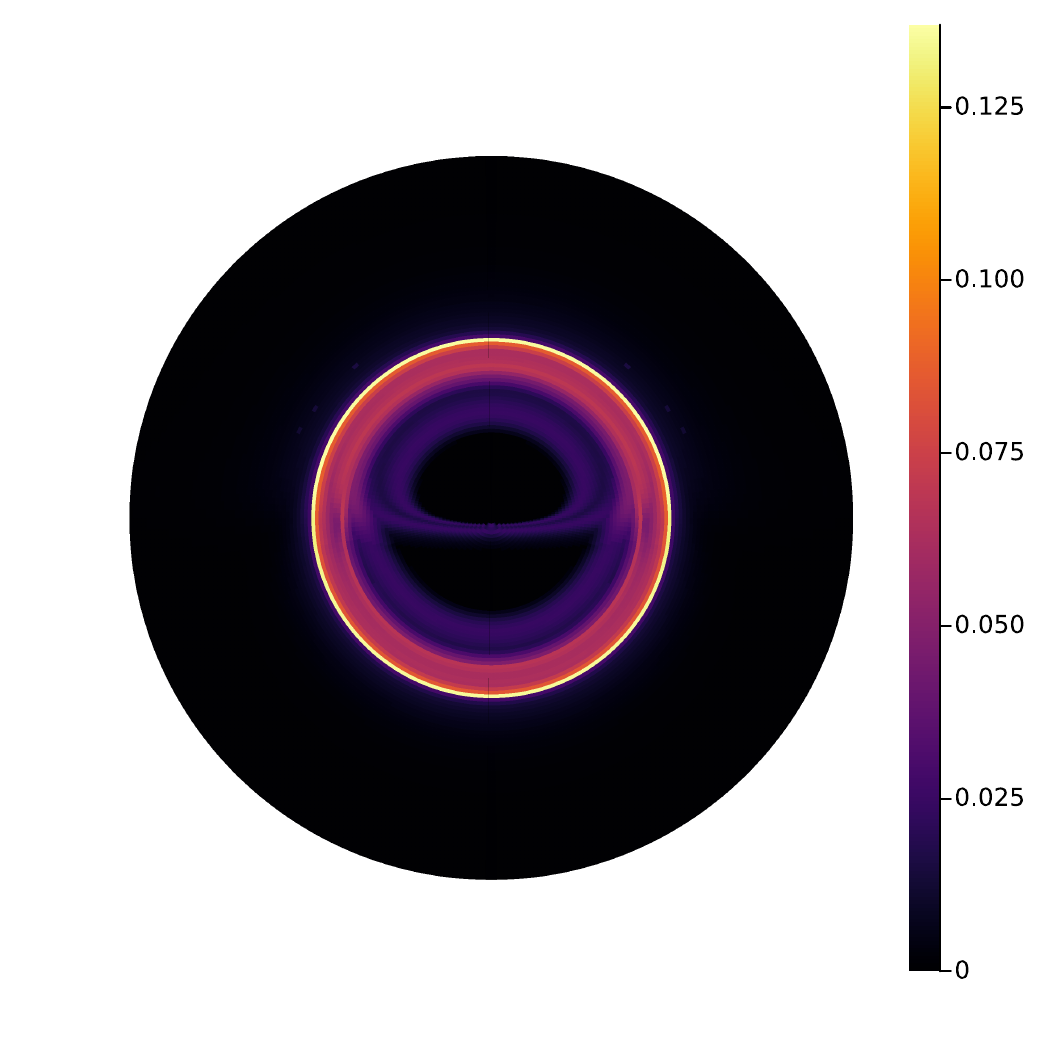} \\
        \includegraphics[width=0.3\textwidth]{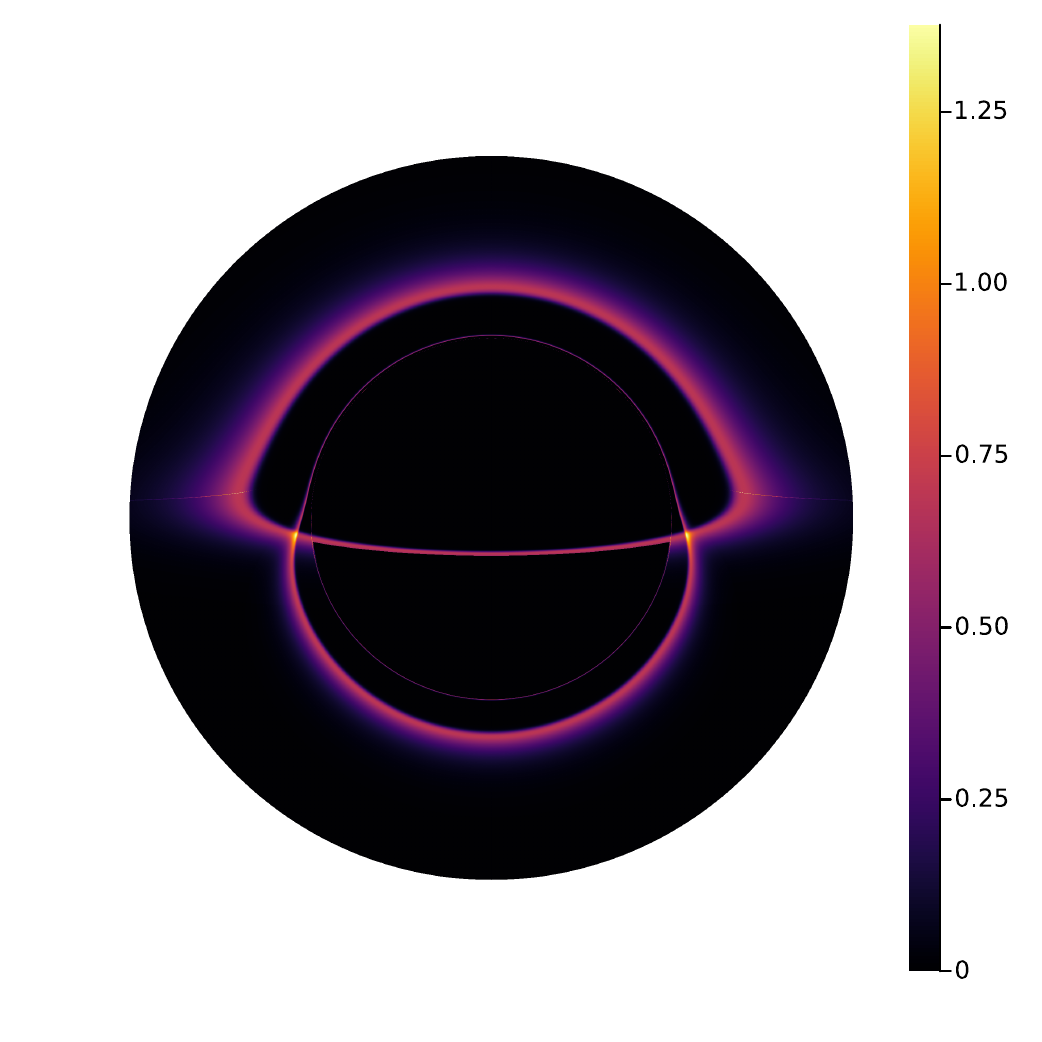} &
        \includegraphics[width=0.3\textwidth]{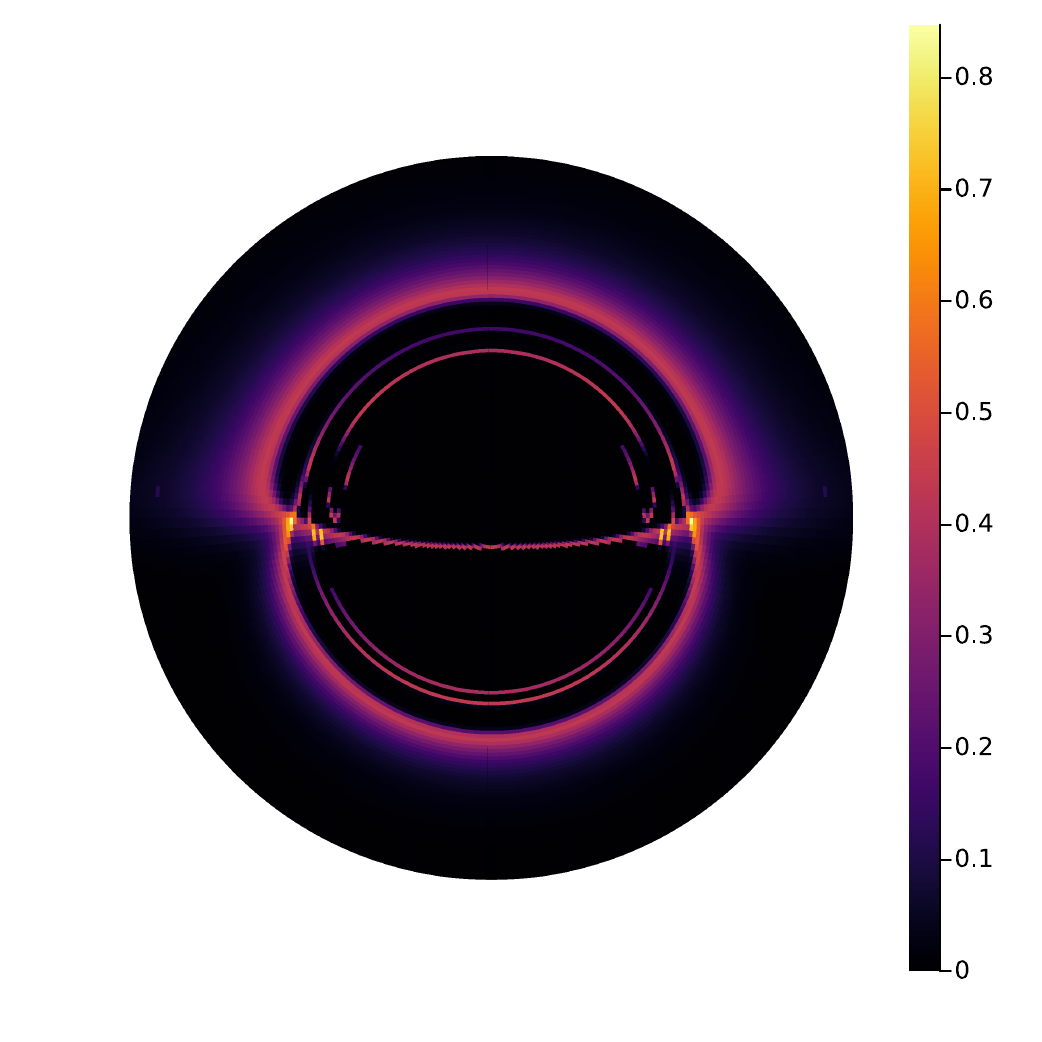} &
        \includegraphics[width=0.3\textwidth]{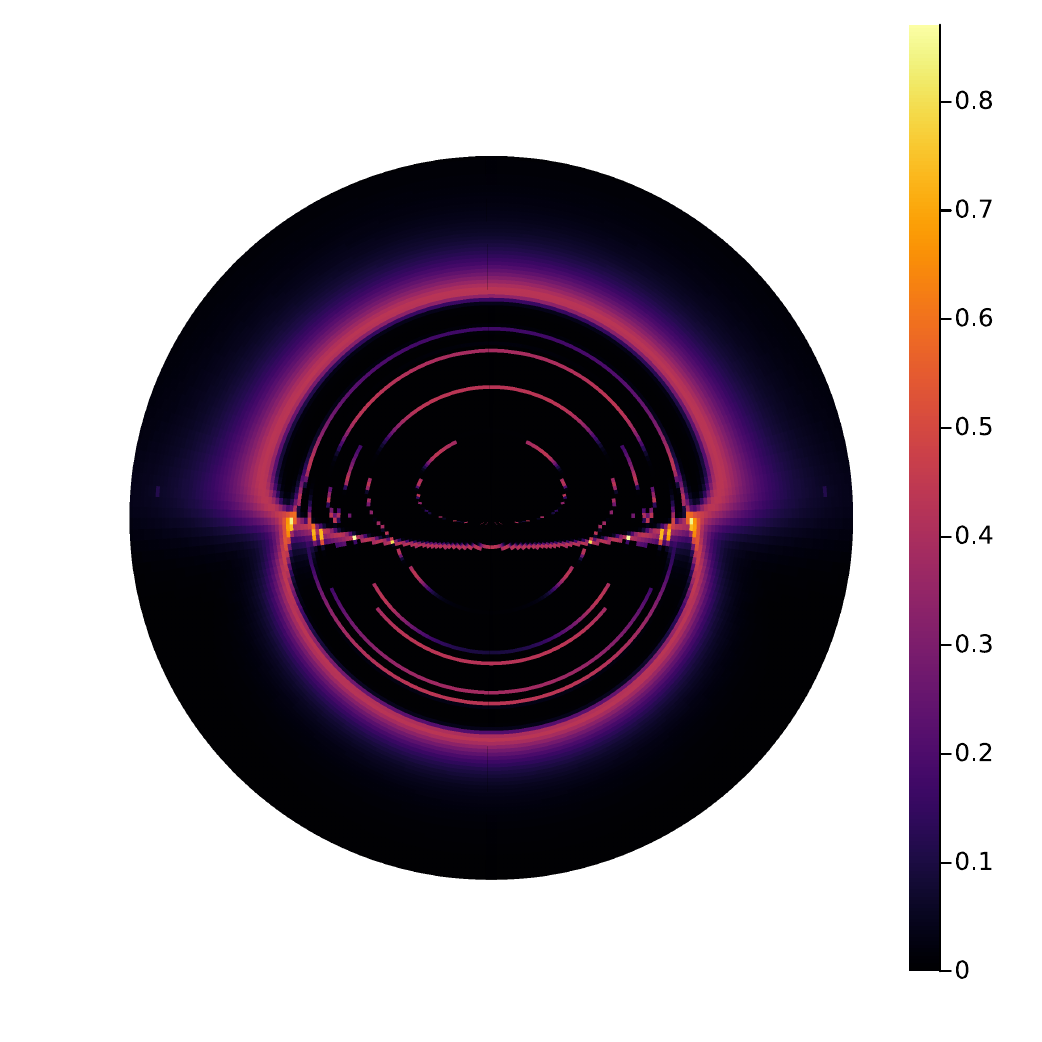} \\
    \end{tabular}
    \caption{Inclined observational images with inclination angle $i = 80^{\circ}$. The left column corresponds to a Reissner-Nordstr\"om black hole, the middle column to an asymmetric wormhole with a single accretion disk, and the right column to two accretion disks. The top row corresponds to the intensity distribution GML1, the middle row to GML2, and the bottom row to GML3.}
    \label{fig:heatmapincl}
\end{figure*}

This setting presents several differences from both the Reissner-Nordstr\"om case and the wormhole with one thin-accretion disk. Indeed, now there are more higher-order photon rings than in the single-disk case and, more importantly, many of them are neatly visible in the optical appearances, either independently or as local insertions of luminosity. This is mainly due to the fact that now each photon ring comes from the potential well the transfer function of Fig. \ref{fig: transfer} displays, which boosts their luminosity. Furthermore, these additional photon rings bring about a strong decrease in the size of the central brightness depression, given the fact that further inner rings appear in the region associated to the well inside the direct emission. In the GLM3 model many peaks appear inner to the direct emission in the observed luminosity, giving rise to six visible photon rings in the full image. On the other hand, for the GML1 and GLM2 models, the wide intensity profile associated with the direct emission now extends further into the inner region, and there are photon rings not only superimposed but also inside it. The former set (already present in the single-disk scenario) boosts the luminosity of the direct emission region, while the latter set is responsible for the reduction in the size of the shadow. 

The bottom line of this discussion is that this two-disk scenario presents both theoretical and observational novelties as compared to canonical black holes and single-disk wormhole scenarios, thus introducing suitable observational discriminators via both photon rings and shadows.

\subsection{Inclined images}

For completeness of our analysis we consider now observations in which there is a non-zero inclination between the observer and the disk\footnote{Note that we neglect the Doppler and relativistic beaming effects that would bring an asymmetric brightness to the image, since this is not the goal pursed here.}, given the fact that this scenario may significantly influence photon ring observables \cite{Salehi:2024cim}. To exaggerate the effects brought by asymmetric wormholes as much as possible, in Fig. \ref{fig:heatmapincl} we depict the images of an inclined observer at a very extreme angle of $80^{\circ}$ (recall that for M87 inclination is estimated to be around $17^{\circ}$) following the same pattern as for the axial images: Reissner-Nordstr\"om black hole (left), asymmetric wormhole with a single disk (middle), and asymmetric wormhole with two disks (right), and the three GLM models (top to bottom). We find similar features for the distribution, visibility, and location of the photon rings as in the axial case, but with some subtle differences. In particular, for the GLM3 model the  morphology of the higher-order rings differs in their shapes near the equatorial plane, while the visibility of some of them improves, particularly in the two-disk case. On the other hand, in the GLM1/GLM2 models the structure of the direct emission region (with some of the higher-order photon rings overlapped on it) now widens and, furthermore, some inner higher-order photon rings become more prominent than in the axial case, particularly for the GLM1 model.

\section{Conclusion and discussion}
\label{C:V}

In this work we have considered the optical appearance of reflection-asymmetric traversable thin-shell wormholes. These wormholes are constructed through a cut-and-paste procedure by which two distinct space-times are matched at a certain hypersurface (the throat) by employing a suitable junction conditions procedure, determining the allowed discontinuities of the corresponding geometrical functions at and across the throat. Such asymmetric wormholes are built within a $f(R)$ extension of GR formulated in the Palatini approach (i.e. with independent metric and affine connection structures) via two Reissner-Nordstr\"om space-times with different masses and charges matched at a hypersurface (the throat), such that the resulting configurations satisfy the energy conditions everywhere and are linearly stable. We take choices for the masses and charges such that the wormhole does not have horizons, while having photon spheres (on both sides of the throat), therefore allowing the bi-directional transit of both matter and energy across the throat.

In this setting, we first investigated the behaviour of individual and bunches of light rays in their transit around and across the wormhole, classified according to the number of half-turns $n$ performed in their trip. We find that the asymmetric nature of the wormhole introduces some new sets of light trajectories that are able to travel through (a portion of) of the other side of wormhole, bringing back information from the geometry there, a possibility which is not available in black hole settings due to the presence of an event horizon. To explore the novelties this possibility could bring in terms of the corresponding optical images, we considered a thin-disk model with a simplified semi-analytic emission profile as given by suitable adaptations of the Standard Unbound distribution, previously employed in the literature to reproduce specific settings of GRMHD simulations. Taking three picks for the coefficients of such profiles (dubbed as GLM models), we built optical images of the asymmetric wormhole in axial and inclined settings for the cases of a single disk on the observer's side, and for the two-disks case (i.e. one additional disk on the other side of the wormhole).

Our asymmetric wormhole scenario reproduces the canonical features of black hole images: a bright ring of radiation generated by the disk's direct emission $n=0$, a set of photon rings (cut in our simulations, for the black hole space-time, at the $n=2$ ring due to their exponentially-decreasing contribution to the luminosity of the image) either stacked with the direct emission or appearing inner to it (depending on the GLM chosen) and a central brightness depression, but it introduces several relevant novelties. In the asymmetric wormhole with the single-disk model (considering trajectories up to $n=7$), several additional visible photon rings are present which, similarly to the usual photon rings, can also be stacked with the direct emission or appear in the innermost region of the image. However, such additional rings are quite faint as compared to the canonical $n=1$ photon ring, largely troubling their observational detectability (similarly to what happens with the canonical $n=2$ ring in black hole images). As opposed to that, in the two-disk case there are further photon rings in the image, which are far more visible than in the single-disk case, and moreover they produce a strong reduction in the shadow's size as compared to both the canonical black hole case and the single-disk wormhole case, a fact that can also act as an observational discriminator of this kind of setting. 

The results found in our work put forward a feature already pointed out in many works in the field: UCOs of different sorts, such as boson stars or wormholes among many others, introduce observational discriminators from canonical black holes under the form of additional higher-order photon rings and a reduction in the size of the central brightness depression. For the observational viability of these additional photon rings, however, there are additional effects that need to be taken into account, including the strong redshift undergone by photons that traverse the region near the throat, which may significantly reduce the luminosity of the additional photon rings and effectively darken the image, spoiling their chances of detectability, as discussed in \cite{Chen:2024ibc}. Furthermore, resolving both the canonical and the new rings is a challenge for VLBI projects \cite{Johnson:2023ynn,Johnson:2024ttr} given its extremely thin character, though ideas to overcome this problem are ongoing \cite{Cardenas-Avendano:2023obg,Keeble:2025gbj,BenAchour:2025uzp}. An exciting possibility that could provide further information on the nature of these rings is given by the quasi-normal models-shadows correspondence recently highlighted in the literature \cite{Pedrotti:2025idg}, whose extension to the UCO case would be interesting to tackle. In particular, the connection between the Lyapunov index of unstable bound geodesics characterizing photon rings and the imaginary part of quasi-normal modes frequency for black holes is worth exploring here, since it could provide further highlights on the structure of  photon rings for devising strategies in their observational detectability. Work along these lines is currently underway. 

\begin{acknowledgments}
This work is supported by the Spanish National
Grants PID2022-138607NBI00 and  CNS2024-154444, funded by MICIU/AEI/10.13039/501100011033 (``ERDF A way of making Europe" and ``PGC
Generaci\'on de Conocimiento") and FEDER, UE. C.F.B.M. acknowledges Coordena\c{c}o de Aperfei\c{c}oamento de Pessoal de N\'ivel Superior – Brasil (CAPES) – Finance Code 001, Fundaç\~ao Amaz\^onia de Amparo a Estudos e Pesquisa (FAPESPA), and Conselho Nacional de Desenvolvimento Cient\'ifico e Tecnol\'ogico (CNPq).
\end{acknowledgments}

\appendix
\renewcommand{\theequation}{A.\arabic{equation}}
\setcounter{equation}{0}

\section*{Appendix}
\addcontentsline{toc}{section}{Apéndice} 

\subsection{Junction conditions for thin-shell wormholes in Palatini $f(R)$ gravity} \label{A:JC}

The junction conditions formalism is the technical support for the cut-and-paste procedure of surgically joining together two patches of space-times at a given hypersurface $\Sigma$ in order to form a single manifold. It is based on the employment of tensorial distributions, i.e., distributions with compact support in the manifold, and its aim is to find the required smoothness of geometrical functions at and across the hypersuface, while at the same time detecting the allowed discontinuities on them without spoiling the well-behaved nature of the field equations everywhere in the full manifold. Originally developed within the framework of GR \cite{Darmois,Israel:1966rt}, in the context of Palatini $f(R)$ gravity such conditions were found in Ref. \cite{Olmo:2020fri} (we refer the reader there for proper definitions and technical aspects) and can be listed as follows (here $+$ and $-$ refer to each side of the manifold, while brackets indicate the discontinuity across the hypersurface:)
\begin{widetext}
\begin{eqnarray}
h_{\mu\nu}^+&=& h_{\mu\nu}^- \quad \text{(Continuity of the induced metric)} \label{JC1} \\
\left[T\right] &=&0 \quad \text{(Continuity of the trace of the energy-momentum tensor)} \label{JC2}\\
\tau &=&0 \quad \text{(Vanishing trace of the energy-momentum tensor at $\Sigma$)}  \label{JC3}\\
-\left[K_{\mu\nu} \right] + \tfrac{1}{3} h_{\mu\nu} \left[K_{\rho}^{\rho} \right] &=& \kappa^2 \frac{\tau_{\mu\nu}}{f_{R,\Sigma}} \quad \text{(Discontinuity of the second fundamental form)} \label{JC4} \\
D^{\rho}\tau_{\rho\nu}&=&-n^{\rho}{h^\sigma}_{\nu} [T_{\rho\sigma}] \quad \text{(First conservation equation)} \label{JC5} \\
(K_{\rho\sigma}^+ + K_{\rho\sigma}^-)\tau^{\rho\sigma} &=&2n^{\rho}n^{\sigma}[T_{\rho\sigma}] -\frac{3R_T^2 f_{RR}^2}{f_R} [b^2] \quad \text{(Second conservation equation)} \label{JC6}
\end{eqnarray}
\end{widetext}
where $T$ is the trace of the energy-momentum tensor $T_{\mu\nu}$, $\tau_{\mu\nu}$ its restriction to the hypersuface $\Sigma$ and $\tau$ its trace, $h_{\mu\nu}=g_{\mu\nu}-n_{\mu}n_{\nu}$ is the first fundamental form with $n^{\mu}$ the unit vector normal to the hypersurface $\Sigma$, $K_{\mu\nu}$ is the second fundamental form, $b\equiv n^{\mu} [\nabla_{\alpha} T]$, and $D^{\rho} \equiv {h^\rho}_{\alpha}\nabla^{\alpha}$ is the covariant derivative on $\Sigma$. These conditions introduce certain novelties as compared to those of GR (and to those of metric $f(R)$ as well). For instance, Eq.(\ref{JC4}) combined with Eq.(\ref{JC2}) imposes different relations between the degrees of freedom driving the behaviour of the matter fields at the hypersurface, which are different from their GR and metric $f(R)$ counterparts \cite{Senovilla:2013vra}, a fact that has important consequences of the sake of building specific applications such as a stellar surfaces or, for the sake of this paper, thin-shell wormholes \cite{Lobo:2020vqh}.  

In fact, in the particular case considered in this work of spherically symmetric space-times with $r=r_0=$constant, the relevant junction conditions to build the solution are those of (\ref{JC4}) and (\ref{JC5}), which read explicitly as
\begin{eqnarray}
[K^\tau_{\ \tau}]- [K^\theta_{\ \theta}]&=&\frac{3\kappa^2}{2f_{R_\Sigma}}\sigma \label{eq:A7} \\
-D_\rho S^{\rho}_{\ \nu}&=&\dot\sigma \ ,
\end{eqnarray}
respectively, besides Eq.(\ref{JC2}). The combination of the last two equations, particularized for a perfect fluid, implies the relation between the pressure and energy density of Eq.(\ref{eq:eos}). This is quite a strong departure from both GR and metric $f(R)$ theories, in that in the Palatini case the matter content on the shell is entirely determined by its energy density, therefore removing any need to fix the equation of state there, as done in the original construction of \cite{Poisson:1995sv}, found within GR. Similarly, Eq.(\ref{eq:A7}) is different from the GR one displayed in \cite{Poisson:1995sv}, in turn imposing different conditions for the positive-definite character and stability of the corresponding thin-shell solutions.

\subsection{Photon orbits and photon rings in spherically symmetric space-times} \label{A:PO}

Consider a spherically symmetric space-time described by the following line element
\begin{equation}
ds^2=-A(r)dt^2+B(r)dr^2+r^2 d\Omega^2,
\end{equation}
where $A(r)$ and $B(r)$ are the two single (radial) functions characterizing it. Without any loss of generality, using the spherical symmetry of the system we can consider only equatorial geodesic motion, $\theta=\pi/2$. Under these conditions, the null geodesic equations read as
\begin{eqnarray}
\dot{t}&=&EA^{-1}, \label{eq:geo1} \\
\dot{\phi}&=& \frac{L}{r^2} \label{eq:geo2}, \\
\dot{r}^2 &=&(AB)^{-1} \left[\frac{1}{b^2}-V(r) \right],  \label{eq:geo3} 
\end{eqnarray}
with $V(r)=\tfrac{A}{r^2}$. The first two equations simply express the conservation of the energy $E$ and angular momentum $L$, as a consequence of the invariance of the system under time reversal and rotations around the azimuthal angle, respectively. Using Eq.(\ref{eq:geo1}) we can rewrite (\ref{eq:geo3}) as
\begin{equation}  \label{eq:geo4}
\frac{d\phi}{dr} = \pm \frac{b}{r^2 \sqrt{1-b^2 V(r)}},
\end{equation}
where the $+$($-$) sign denotes outgoing (ingoing) geodesics. The change in sign in the above equation is given by the turning point $r_t$, such that $\dot{r} \vert_{r=r_t}=0$, that is, $\tfrac{1}{b^2}=V(r_t)$. 

The image in the observer's plane corresponds to the lensed trajectories of photons due to gravitational deflection, and it is generated out of a ray-tracing procedure: the backtracking of every light ray that appears on the observer's plane image by using Eq.(\ref{eq:geo4}) in order to determine the location at which it was generated. In this procedure, the \textit{photon sphere}, namely, the surface of unstable bound geodesics, plays a key role. This surface corresponds to the critical (maxima) points of the effective potential, namely
\begin{equation}
\frac{1}{b_c}=V(r_{ps}) ; \quad \frac{dV}{dr}\Big\vert_{r=r_{ps}} =0 ; \quad \frac{d^2V}{dr^2} \Big\vert_{r=r_{ps}}<0,
\end{equation}
with the last equation defining the maxima character of the surface (the opposite case, i.e., minima of the potential, it is called anti-photon sphere instead, and both black holes and UCOs are capable of holding both \cite{Guo:2022ghl}). These equations also define the critical impact parameter $b_c$, i.e., the impact parameter needed for a photon to asymptotically approach the photon sphere $r_{ps}$.  These definitions help to classify light trajectories; in a typical black hole space-time, photons with $b>b_c$ are scattered by the black hole, while those with $b<b_c$ are swallowed by its event horizon. Those with $b \gtrsim b_c$ produce highly-bent trajectories that can turn several times around the black hole and which appear, in the observer's plane image, as a series of photon rings indexed by the number $n$ of half-turns performed around the black hole and which closely track, in the limit $n \to \infty$, a critical curve there (corresponding to the projection of the photon sphere).

\end{document}